\documentclass[hidelinks,11pt,letterpaper]{article}
\usepackage[sectionbib,square]{natbib}
\usepackage[margin=1in]{geometry}
\usepackage{inconsolata}
\usepackage[T1]{fontenc}
\usepackage[utf8]{inputenc}
\usepackage{setspace}
\usepackage{microtype}
\usepackage{amsfonts,amsmath,amssymb,amsthm}
\usepackage{fontawesome}
\usepackage{booktabs}
\usepackage{subcaption}
\usepackage{wrapfig}
\usepackage{multicol}
\usepackage{pifont}
\usepackage[usenames,dvipsnames]{xcolor}
\usepackage[font={small,color=black,onehalfspacing}, labelfont={bf}]{caption}
\usepackage{graphicx}
\usepackage[export]{adjustbox}
\usepackage{csquotes}
\usepackage{xurl}
\usepackage{mwe}
\usepackage{enumitem}
\usepackage{titlesec}
\usepackage{palatino}
\usepackage{lipsum}
\usepackage{hyperref}       %
\usepackage{nicefrac}       %
\usepackage[english]{babel}
\usepackage{multirow}

\newcommand{\cmark}{\ding{51}}%
\newcommand{\xmark}{\ding{55}}%

\newtheorem{assumption}{Assumption}

\titleformat{\subsubsection}[runin]
       {\normalfont\bfseries}
       {\thesubsubsection}
       {0.5em}
       {}
       [.]
\titlespacing*{\subsubsection}{\parindent}{*0}{0.25em}

\newenvironment{sciabstract}{%
\begin{quote}\small\onehalfspacing}
{\end{quote}}

\title{Designing Effective Music Excerpts
}

\author
{
  Emaad Manzoor$^{*}$\\
  Cornell University\\
  \and
  Nikhil Malik\\
  University of Southern California\\
}

\date{}

\begin{document}

\maketitle
\thispagestyle{empty}

\onehalfspacing

\begin{center}
\textbf{Abstract}
\end{center}
\begin{sciabstract}
  Excerpts are widely used to preview and promote musical works. Effective excerpts induce consumption of the source musical work and thus generate revenue.
  Yet, what makes an excerpt effective remains unexplored.
  We leverage a policy change by Apple that generates quasi-exogenous variation in the excerpts of songs in the iTunes Music Store to estimate that having a 60 second longer excerpt increases songs' unique monthly listeners by 5.4\% on average, by 9.7\% for lesser known songs, and by 11.1\% for lesser known artists. This is comparable to the impact of being featured on the Spotify Global Top 50 playlist.
  We develop measures of musical repetition and unpredictability to examine information provision as a mechanism, and find that the demand-enhancing effect of longer excerpts is suppressed when they are repetitive, too predictable, or too unpredictable. Our findings support platforms' adoption of longer excerpts to improve content discovery and our measures can help inform excerpt selection in practice.
  
\end{sciabstract}

\begin{quote}\small\textbf{Keywords:} digital media products, free trials, music consumption, generative machine learning \end{quote}

{\let\thefootnote\relax\footnotetext{$^{*}$Address correspondence to: \underline{\smash{\href{mailto:emaadmanzoor@cornell.edu}{emaadmanzoor@cornell.edu}}}. We thank Kwabena B. Donkor, Sachin Gupta, Sylvia Hristakeva, Shreya Kankanhalli, and Alan L. Montgomery for comments that improved the paper. We are also grateful to J$\bar{\textrm{u}}$ra Liaukonyt$\dot{\textrm{e}}$ for facilitating access to the U.S. music sales data, to Rachee Singh for lending us Nvidia A100 GPUs, and to Michael D. Smith and Joshua Friedlander for their insights into the recorded music industry. Supporting data and software will be made available at: \underline{\smash{\href{http://emaadmanzoor.com/effective-music-excerpts}{http://emaadmanzoor.com/effective-music-excerpts/}}}.}}

\clearpage

\doublespacing

\pagenumbering{arabic}

\section{Introduction}
\label{sec:introduction}


Excerpts, such as animated thumbnails on YouTube and music clips on TikTok, serve as a ``lubricant in the market for creative content'' \citep{toubia2021poisson}.
The rapid growth of short-form social media has spurred creators' adoption of excerpts as a promotional tool \citep{bi2023},
where their impact is largely determined by how well they induce long-form content consumption\footnote{While TikTok now pays royalties, Spotify's streaming royalties are 500 times higher on average \citep{bb2022_2}.
However, TikTok views seldom translate into Spotify streams and listeners, even for viral TikTok content \citep{pudding2022}.
}.
Excerpts are also used by media platforms to facilitate content sharing and discovery.
Creators and platforms are thus challenged to design excerpts that drive consumption of the full content.

Excerpt length is a key design decision\footnote{TikTok, Instagram Reels, and YouTube Shorts now allow videos up to 60, 90, and 60 seconds long, respectively.
}.
Longer excerpts can increase demand by providing consumers more information, but can also cannibalize demand if consumers derive negligible additional utility from the full content (eg. if the excerpt is a near-substitute). Further, consumers can decide not to consume the excerpt beyond a length threshold (eg. due to time constraints), in which case increasing excerpt length beyond that threshold has no effect. Whether and how excerpt length matters for media products is thus theoretically ambiguous and an empirical question.

In this work, we focus on excerpt design for music --- and industry that contributes \$170 billion annually to the U.S. GDP \citep{riaa2023-2} --- motivated by its central role in contemporary social media \citep{radovanovic2022tiktok}. Music is a quintessential experience good, of uncertain value until consumption. Effective excerpts can incentivize consumers to explore, who might otherwise default to popular content that is easier to value. Incomplete demand estimates generated by uninformed consumers also motivate creators and financiers to overinvest in what's popular. Effective excerpts can lower the cost of exploration and potentially increase investment in emerging content.

While prior research has not examined music excerpts specifically, the choice of excerpt length resembles that of quantity or duration in free trials of information goods \citep{bawa2004effects,chellappa2005managing}. The nascent empirical research on this choice recommends shorter software trials \citep{yoganarasimhan2023design} and fewer free news articles \citep{aral2021digital} to increase subscriptions. Extrapolating these recommendations to music is difficult, since the short experiential duration, large content variety, and hedonicity of music differentiates how consumers value music from how they value news and software subscriptions.

\clearpage

While music better exemplifies digital media products such as movies and video games, prior research on excerpts of media products treats the excerpt length as fixed \citep{liu2018video,li2019optimal}. Fixing the excerpt length limits an important driver of cannibalization. By studying the impact of excerpt length on demand net of any cannibalization, we seek to quantify the overall tradeoff between cannibalization and information provision for digital media products.

To quantify how excerpts impact music consumption, we leverage quasi-exogenous variation generated by a policy change on the Apple iTunes Music Store that increased the lengths of free previews of songs longer than a threshold from 30 seconds to 90 seconds. This policy change occurred during a time when the iTunes Music Store accounted for 63-70\% of all U.S. digital music sales, which in turn accounted for 40.9\% of all U.S. recorded music revenue \citep{riaa2022}.

Digital music sales (i.e. paid downloads) are a particularly compelling signal of consumer choice. Paid downloads are ``active'' choices that (arguably) better reflect consumer preferences than the passive listening observed in streaming music consumption. In fact, for charting (eg. on Billboard) and sales certifications (eg. RIAA Gold and Platinum awards), the music industry equates each paid download to 150 streams \citep{nielsen2017,riaa2023}.


To leverage the iTunes policy change, we require both the pre-policy and the post-policy preview of each song. The pre-policy previews, however, are no longer visible in the iTunes Music Store. To address this challenge, we locate a novel data source --- Apple's Enterprise Partner Feed \citep{apple2023} --- containing the pre-policy previews of \textit{all} songs in the iTunes Music Store. We combine this data with music consumption data to analyze the pre- and post-policy sales of songs with excerpts affected by the policy change relative to those unaffected by the policy change.

We estimate that longer excerpts increase monthly sales by 5.4\%, and by 9.7\% for less popular songs.
Since each sale originates from a unique iTunes account, our estimate can be interpreted as an increase in the number of unique monthly listeners of a song. Longer excerpts thus increase music consumption at least as much as being included in the Spotify Global Top 50 playlist\footnote{\cite{aguiar2018platforms} estimate that ``being on the Global Top 50 list raises a song's
streams by about 3 million, or by about 3.3 percent of the average streams for songs that make the Global Top 50''.
},
without the costs of acquiring sufficient streams for playlist inclusion.
To evaluate the existence of confounders coincident with the timing of the policy change, we estimate but find no evidence of the policy change impacting \textit{non}-iTunes music consumption via radio and online streaming.


\textbf{Mechanisms.}~~The information provided by an excerpt is a function of its content, and not just its length. In our context, the pre-policy audio of an excerpt that is affected by the policy change differs from its post-policy audio along dimensions apart from length. We leverage this unique excerpt-level variation in content to study information provision as a mechanism. Specifically, we develop two measures of information in excerpts' audio --- musical repetition and unpredictability --- and examine how these measures moderate the impact of longer excerpts on demand.


\textit{Repetition.}~~Intuitively, repetition in an excerpt of a media product reduces the amount of distinct information available for consumers to learn from. Hence, if information provision is an underlying mechanism, increasing repetition should suppress the demand-enhancing effect of longer excerpts.

Our measure of musical repetition is based on digital compression algorithms.
Compression algorithms identify repetition to encode data using fewer bits. For example, using the run-length compression algorithm \citep{robinson1967results}, the 10-byte sequence ``aaaaabbbbb'' (assuming 1 byte per character) is compressed into the 4-byte encoded sequence ``5a5b''. In contrast, the less-repetitive 10-byte sequence ``aabbabbaba'' is compressed into the 10-byte encoded sequence ``2a2ba2baba''.
More sophisticated algorithms such as Lempel-Ziv-Welch \citep{ziv1978compression,welch1984technique} and audio-specific algorithms such as MP3 and FLAC \citep{salomon2006compression} produce shorter encoded sequences at a higher computational cost.

In general, after compression, repetitive sequences have shorter encoded lengths. Hence, we quantify repetition in an excerpt using the encoded length
of its compressed audio (lower encoded lengths imply higher repetition). 
We find that the demand-enhancing effect of longer excerpts is suppressed when they are repetitive. Specifically, increasing the encoded length (i.e. decreasing repetition) from the first to the last decile is associated with a 6 percentage point increase in the demand-enhancing effect of a 90-second excerpt, supporting the information provision mechanism.


\textit{Unpredictability.}~~Prior research finds that music consumers attempt to predict the upcoming musical notes in a song as it unfolds over time \citep{meyer2008emotion,huron2008sweet,pearce2012auditory}. Based on this behavior, we expect musical predictability in excerpts to also reduce information, over and above repetition. Our intuition is that predictable musical notes (i.e. notes that the consumer expects, based on their priors) provide little marginal information even when the notes are distinct.
In the hypothetical extreme case, an excerpt that a consumer has recently listened to is perfectly predictable, and is thus uninformative even if it is not repetitive.

To measure musical unpredictability, we leverage ``artificially intelligent'' music consumers represented by generative machine learning models of music \citep{dhariwal2020jukebox,copet2023simple}. These models are trained on tens of thousands of hours of music 
to generate music both unconditionally, and conditionally as a continuation of an audio clip.
Various regularization techniques used during training ensure that these models generalize creatively out-of-sample, instead of simply memorizing the music they are trained on. 

We quantify the musical unpredictability of an excerpt as the log-perplexity \citep{chen1998evaluation} assigned to it by an autoregressive
generative music model.
Intuitively, the more unexpected a generative model finds a sequence on average, the higher its log-perplexity. For musical audio in particular, the log-perplexity can be viewed as the \textit{in}accuracy of a generative music model in predicting the continuation of the audio from a portion of the audio, averaged over all portions.

We find that the demand-enhancing effect of longer excerpts is suppressed when they are too predictable, consistent with information provision as a mechanism. However, we also find that the demand-enhancing effect of longer excerpts is suppressed when they are too unpredictable. Behaviorally, the ``inverted U-shape'' of this moderation effect can be explained by a psychobiological liking for moderate levels of musical surprise\footnote{
\cite{pearce2012auditory} suggest that ``intermediate degrees of predictability are preferred with very predictable and very unpredictable music (with respect to prior knowledge) both being disliked''.} \citep{north1995subjective,pearce2012auditory}.
We can also explain this finding with a simple model, if extreme unpredictability diminishes consumers' ability to develop valuations of the full song.
\textbf{Contributions.} We make three main contributions. Substantively, we provide the first estimates of the impact of excerpt length on music consumption. We thus extend the literature on free trial design \citep{yoganarasimhan2023design,li2019optimal,aral2021digital}, and contribute to the broader literature on media products \citep{eliashberg2006motion,boughanmi2021dynamics,simonov2023suspense}.
Methodologically, we measure information directly from the excerpt audio to examine the information provision mechanism. Our measures are adaptable to other modalities such as images and video, and complement the marketing literature on methodology for audiovisual data \citep{rajaram2020video,fong2021theory,yang2022high,yang2023first}. 
Managerially, we provide creators and marketers guiding measures and empirical evidence of the capabilities and constraints of excerpts, which are also usable in audio advertisements \citep{youtube2023,spotifyads2023}.


TikTok has emerged as one of the largest music discovery platforms today \citep{cnbc2023}. For the billions of users on platforms like TikTok, their first introduction to a musical work often takes the form of an excerpt. Our findings indicate that short-form social media platforms' adoption of longer excerpts can improve consumer-content matching efficiency, benefiting their ecosystems overall. However, we also expect such platforms to experience reduced engagement as consumers go off-platform to consume long-form versions of their discoveries. This rationalizes TikTok's launch of the TikTok Music streaming service to vertically integrate long-form content distribution.

We proceed in 6 sections after discussing related work in Section \ref{sec:related}. In Section \ref{sec:theory}, we present a simple model to motivate our analyses. In Sections \ref{sec:natural} and \ref{sec:data}, we discuss our institutional background and dataset. In Section \ref{sec:policy}, we quantify the impact of the iTunes policy change and analyze effect heterogeneity. In Section \ref{sec:information}, we examine information provision as a mechanism. Section \ref{sec:conclusion} concludes.


\section{Related Work}
\label{sec:related}

Our work is related to research on the optimal quantity and duration of free trials. \cite{yoganarasimhan2023design} study the impact of trial duration in markets for software, and find that shorter trials produce higher subscription rates than longer trials. Similarly, \cite{aral2021digital} analyze the impact of the quantity and diversity of free news articles on subscriptions, and find that subscription rates increase with fewer and more diverse free articles. 
Valuing hedonic products such as music is often more difficult than valuing the functional benefits of goods such as software and online news subscriptions \citep{okada2005justification}. Hence, shorter excerpts may not be optimal for media products if consumers require more information for their valuation, motivating our work.


Our work is related to the literature on information and product discovery in cultural markets. \cite{hendricks2009information} find that backward spillovers from new album releases increase older albums' sales.
\cite{zhang2018intellectual} show that relaxing digital music sharing restrictions by increases digital music sales.
\cite{ryoo2021spoilers} show that spoilers in reviews increase box office revenue. Extensive research studies piracy as an information provision mechanism for digital media products \citep{chellappa2005managing,lu2020does}.
Our work complements this literature by quantifying the impact of excerpts as an information provision mechanism, that is readily usable by emerging artists, and increasingly applicable with the growth of short-form social media.

In the behavioral literature, our research is related to \citep{nunes2015power}, who study the impact of lyrical repetition on the consumption of pop music. \cite{nunes2015power} find that repetition increases processing fluency and is associated with higher consumption.
Our work is focused on consumers' decisions
to listen to the full song based on its excerpt, and not on consumers' preferences for the excerpt itself. Hence, we focus on the potential information-reducing effects of repetition, instead of its consumption-enhancing effect on excerpts. Further, we measure repetition in audio, which consumers may process differently than lyrical or lexical repetition.

Methodologically, our research is related to the literature on audiovisual analytics \citep{wang2021audio,peter2023artificial,chakraborty2023ai,rajaram2020video,fong2021theory,yang2022high,yang2023first}.
Our measure of unpredictability is related to suspense and surprise \citep{liu2020passive,simonov2023suspense}, but does not require viewership data to calculate.
Our measure of unpredictability is also related to research that approximates consumers with generative machine learning models of language \citep{horton2023large,goli2023crosslinguistic,grossmann2023ai,brand2023using,mills2023opportunities,qiu2023much}.

More generally, our work is related to the literature on fine arts products \citep{kim2020valuing}, and to the literature on music consumption \citep{bradlow2001bayesian,datta2018changing,boughanmi2021dynamics,boughanmi2022contextual}. While prior research on music consumption focuses on attributes of and the demand for full songs, our work focuses on attributes of short music excerpts. Specifically, we focus on the informational capabilities of short music excerpts, and not on the demand for the excerpts themselves. Though listening to an excerpt (and how much of it to listen to) is itself a choice, we delegate studying this choice to future work.

\section{Theoretical Background}
\label{sec:theory}

Extensive theoretical research has examined the tradeoff between information provision and demand cannibalization in free trials of information goods \citep{wang2023endogenous,li2019optimal,halbheer2014choosing,xiang2011preview,wang2009sampling}. We distill the key aspects of this literature into a simple model of music consumption to motivate our empirical analyses.

\clearpage

Consider a consumer evaluating their decision to listen to a song $j$. We denote by $m_{j} \in \textrm{Uniform}(0,1)$ the match value of song $j$ for the consumer. We model a song's excerpt as providing consumers a noisy signal of the match value. We define a consumer's perceived match value as a convex combination of their prior $p \in (0,1)$ and the noisy signal provided by the excerpt, with a lower weight $\theta \in (0,1)$ assigned by consumers to less informative signals:
\begin{align}
  \textrm{Perceived match value}_{j} &= (1-\theta) p + \theta m_{j}
\end{align}
We model listening decisions as based on the perceived match value exceeding a threshold $\tau$ (reflecting the costs consumers incur in terms of effort or time). Note that $\tau > p$ is necessarily true since, if $\tau < p$, consumers would (unrealistically) buy every song when $\theta=0$. The demand for a song from a unit mass of consumers (ignoring cannibalization for now) is given by:
\begin{align}
  \label{eq:theoreticaldemand}
  \textrm{Demand } D_{j}
  &=1 - \dfrac{\tau - (1 - \theta) p}{\theta}, \qquad \textrm{ if } \tau \in (p, \theta + (1-\theta)p)\nonumber\\
  &\implies
  \dfrac{\textit{d}D_{j}}{\textit{d}\theta} = \dfrac{\tau - p}{\theta^2} > 0, \nonumber\\
  &~~~\qquad \dfrac{\textit{d}^2 D_{j}}{\textit{d}\theta\textit{d}p} = -\dfrac{1}{\theta^2} < 0
\end{align}
Equation \ref{eq:theoreticaldemand} shows that in this simple model, when the consumption cost $\tau$ is not too high, demand increases with the excerpt informativeness $\theta$. Equation \ref{eq:theoreticaldemand} also shows that the rate of change of demand with $\theta$ decreases with increasing $p$. Intuitively, the demand-enhancing effect of informative excerpts is suppressed when consumers are better informed (eg. for popular songs).

Excerpt informativeness can be increased by increasing its length. However, this decreases the song portion \textit{not} available for free, potentially reducing demand due to self cannibalization. The cannibalization risk can be lowered by frictions such as the inability to listen to excerpts via external media devices or to add excerpts to personal playlists. Hence, to what extent market expansion dominates cannibalization is an empirical question that motivates our analyses in Section \ref{sec:policy}.

Further, excerpt informativeness is a function of its content, and not just its length. In Section \ref{sec:information}, we measure two information-reducing properties of excerpt content: repetition and predictability. Based on Equation \ref{eq:theoreticaldemand}, we expect the impact of longer excerpts to be suppressed when their informativeness is reduced by increasing repetition and predictability. 

\clearpage

\begin{figure*}[!t]
  \centering
  \caption{U.S. digital single sales revenue (based on public data from \citep{riaa2022})}
  \includegraphics[width=\linewidth]{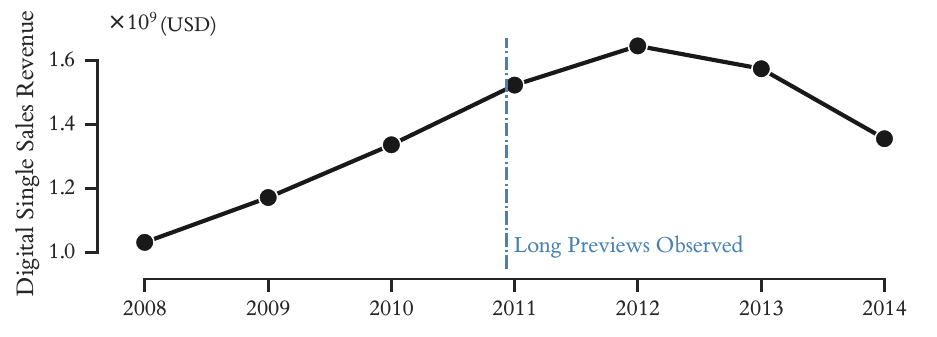}
  \caption*{\footnotesize\textit{Notes:} On November 2, 2010, Apple notified record labels that the preview lengths of tracks at least 2.5 minutes long on the iTunes Music Store would increase from 30 to 90 seconds \citep{macrumors2010}. Longer previews were first observed in the U.S. iTunes Store on December 8, 2010 (dashed blue line) \citep{macrumors2010-4}, after delays due to legal objections \citep{macrumors2010-3}.}
  \label{fig:industrytrends}
\end{figure*}

\section{The Natural Experiment}
\label{sec:natural}

\subsection{Background} The iTunes Music Store was launched by Apple in 2003, offering consumers paid digital downloads of albums and individual tracks. Since January 2009, track pricing is almost-uniform with three possible prices (set by record labels): \$0.69, \$0.99, or \$1.29.
Between 2010 and 2012, the iTunes Music Store accounted for 63-70\% of all U.S. digital music sales, with the Amazon MP3 Store having the next largest market share of 12-22\% \citep{reuters2010,reuters2013}. At its peak in 2012, paid music downloads accounted for 40.9\% of all U.S. recorded music revenue \citep{riaa2022}.

Despite having been overtaken by music streaming in 2015, music downloads are still a valued signal of consumer choice today. The Billboard charts and the RIAA (which awards Gold and Platinum certifications) equate 1 download to 150 on-demand streams \citep{nielsen2017,riaa2023}.
Figure \ref{fig:industrytrends} shows the total U.S. digital single sales revenue\footnote{We report revenue numbers from \citep{riaa2022} to maintain the confidentiality of our music consumption data.} in each year between 2008 and 2014.
In our empirical analyses, we focus on digital single sales between 2010 and 2012, coinciding with the peak of the downloading era.

\subsection{The iTunes Preview Policy Change}

On November 2, 2010, Apple announced that they would soon be increasing the preview lengths of tracks on the iTunes Music Store that were at least 2.5 minutes long from 30 seconds to 90 seconds \citep{macrumors2010}. The full announcement is in Appendix A.

This announcement was expected at the Apple special event in September 2010.
However, last-minute legal objections by the National Music Publishers Association delayed the policy enactment till December 8, 2010 \citep{macrumors2010-3}, when longer previews were first observed in the U.S. iTunes Music Store \citep{macrumors2010-4}. These events suggest that the exact timing of the policy change was driven in part by unanticipated legal hurdles.

The Amazon MP3 Store, Napster, Rhapsody, and non-U.S. iTunes storefronts until July 2011 \citep{macrumors2011} continued to offer 30 second previews. 
Apart from free previews on digital music stores, consumers could also sample music on programmed and on-demand streaming platforms such as Pandora and YouTube, and via cable television, satellite radio, and terrestrial radio.



We leverage the iTunes preview policy change as a natural experiment. Specifically, we analyze changes in the trends of digital sales of tracks affected by the policy change relative to tracks unaffected by the policy change in a difference-in-differences framework.

\section{Data}
\label{sec:data}

We construct a dataset by merging the iTunes Music Store catalog with monthly U.S. music consumption data between March 1, 2010 and August 31, 2011. This 18-month time period is centered around December 2010, the month in which longer iTunes previews were first observed. In this section, we detail our data sources and dataset construction process.

\subsection{U.S. Music Consumption Data}
\label{sec:luminate}

We obtain U.S. music consumption data from Luminate (formerly Nielsen SoundScan), which is the industry standard for recorded music consumption tracking in the United States. Luminate measures digital music consumption in online music stores and on streaming platforms, physical (album) music consumption, and music consumption via satellite and terrestrial radio.

\subsubsection{Digital single sales} Luminate reports digital single sales at the level of \textit{recordings}. A recording is a specific instance of a musical performance after it has been mixed and transferred to a physical or digital storage medium. Each recording is identified by an International Standard Reporting Code (ISRC). Recordings are used to produce one or more \textit{tracks} that are made available for purchase in online stores. Tracks produced from the same recording are aurally indistinguishable (apart from silent sections). The digital single sales reported by Luminate for each recording are the sum of the sales of all the tracks produced from that recording. Along with the sales, Luminate also reports each recording's first release date, core and secondary genres, and record label.

Luminate aggregates recording sales over all online music stores. Since the iTunes Music Store accounted for 63-70\% of all U.S. digital music sales between 2010 and 2012, we expect that iTunes sales comprise the majority of digital single sales reported by Luminate during this period. The Amazon MP3 store, with the next largest market share of 12-22\%, did not change their preview length from 30 seconds between 2010 and 2012. Hence, misattributing sales on the Amazon MP3 Store to iTunes would produce conservative estimates of the impact of the iTunes preview policy change (i.e. biased towards finding no effect).

\subsubsection{Streams and radio airplay audience impressions} Luminate reports on-demand streams on interactive streaming platforms such as YouTube, Rhapsody, and Spotify, programmed streams on non-interactive streaming platforms such as Pandora, and radio airplay audience impressions at the \textit{artist-title} level. Analogous to digital single sales, stream counts are aggregated for each artist-title across streaming platforms.

\subsubsection{Ranked lists and aggregated reports} For the digital single sales, streams, and radio airplay consumption metrics, Luminate requires a list of artist-titles as input. Luminate also provides ranked lists of artist-titles (limited to the top 10,000 artist-titles in each aggregated report) and industry-wide consumption trends for any given consumption metric, time period, and musical genre. These aggregated reports do not require a list of artist-titles as input.

\subsubsection{Relating recordings, tracks, and artist-titles} To clarify the relationship between recordings, tracks, and artist-titles, consider for example the recording with ISRC \textrm{GBAYE9200070}, titled ``Creep'' by the artist Radiohead. This recording was used to produce over a 100 tracks on various physical and digital albums between 1992 and 2021 \citep{mb2023-1}. The track durations vary between 233 and 240 seconds due to silence (used to create inter-track gaps on albums) added to the recording's start or end. The radio edit of this recording has a different ISRC \textrm{GBAYE9200349} that was used to produce several tracks with the same artist-title as the original recording. Overall, there are over 50 recordings titled ``Creep'' by Radiohead representing other edits and live performances.

\subsection{iTunes Music Store Catalog Data}
\label{sec:epf}

We obtain a snapshot of the entire iTunes Music Store catalog (as of May 2023) via Apple's Enterprise Partner Feed \citep{apple2023}. We retain only tracks that are available for purchase in the United States. Our snapshot contains a variety of metadata, from which we retrieve each track's duration and recording ISRC. Note that tracks are added to and removed from the catalog each week, and some tracks that were available for purchase between 2010 and 2012 may be missing in our snapshot.

Our snapshot contains a link to the webpage of each track in the iTunes Music Store, which contains the track's consumer-facing post-policy previews. Additionally, the snapshot contains a link to the pre-policy preview of each track \textit{including} tracks longer than 150 seconds. These previews are no longer exposed to consumers in the iTunes Music Store. We use these two links to retrieve the pre-policy \textit{and} post-policy previews of each track for subsequent analyses.

\subsection{Implications of Taste Depreciation}
\label{sec:depreciation}

Music consumption is characterized by rapid taste depreciation. \cite{garcia2020copyright} find that digital single sales decay to 20\% of their peak after a year post-release on average, and \cite{fix2022} finds that the median time taken by a Spotify Top 200 hit to decay to half its peak stream count is 59 days. Taste depreciation poses a challenge for difference-in-differences analyses when consumption is analyzed near the end of the depreciation curve. We discuss this issue with a simulated example. This issue motivates aspects of our data construction strategy detailed in Section \ref{sec:construction}.

\begin{figure*}[!t]
  \centering
  \caption{Simulated monthly consumption trends for a hypothetical treated and control recording}
  \begin{subfigure}[t]{0.49\textwidth}
    \centering
    \includegraphics[width=\textwidth]{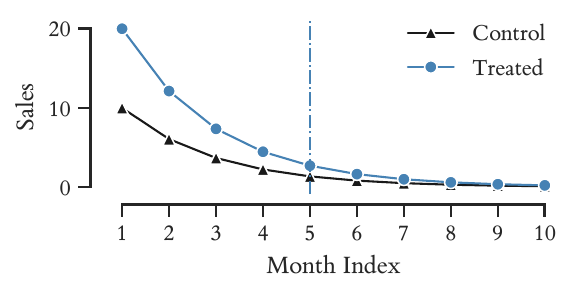}
    \caption{Monthly sales}
  \end{subfigure}
  \begin{subfigure}[t]{0.49\textwidth}
    \centering
    \includegraphics[width=\textwidth]{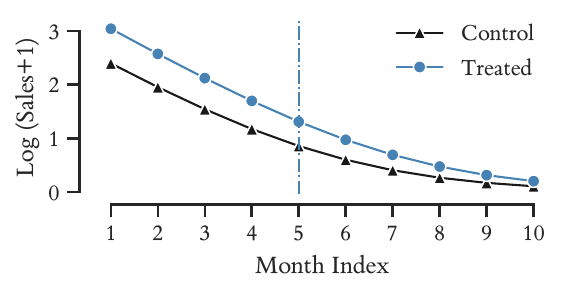}
    \caption{Monthly log sales}
  \end{subfigure} 
  \caption*{\footnotesize\textit{Notes:} Figure \ref{fig:simulation} plots the (a) monthly sales, and (b) monthly log sales (incremented by 1) of 2 hypothetical recordings: one treated and one control. The vertical blue line indicates a hypothetical "policy change month" for a policy with no effect.} 
  \label{fig:simulation}
\end{figure*}
\subsubsection{Simulated example} Consider two hypothetical recordings, one treated and one control, that are released simultaneously and decay exponentially at the same rate. Their simulated\footnote{We set $\textrm{Sales}_t = Ae^{-t/2}$ where $A=20$ for the treated and $A=10$ for the control recording.} sales and log sales (incremented by 1) are plotted in Figure \ref{fig:simulation}, where the vertical blue line indicates the "policy change month" for a policy with no effect. Since the decay rates of the recordings are identical, their log sales are roughly\footnote{The log sales are not exactly parallel because we increment the sales by 1 before taking their logarithm.} parallel before the policy change. However, their log sales converge to zero after the policy change.
This example reveals that analyzing consumption near the end of products' taste depreciation curves mechanically violates the parallel trends assumption required to interpret difference-in-differences estimates as causal effects.

\subsubsection{Implications for dataset construction} The aforementioned example motivates two aspects of our dataset construction strategy. First, we only consider artist-titles in the rock genre. Rock music has a flatter depreciation curve than other genres \citep{garcia2020copyright}. In 2010 in particular, 72.4\% of rock music sales (and only 33.9\% of rap music sales, in contrast) were of music released during or before 2008 \citep{billboard2011}. Rock music sales also comprised the majority of digital single sales in 2010 \citep{billboard2011}. Second, we only consider the top selling artist-titles in terms of digital single sales in the years immediately prior to the policy change, with the expectation that these artist-titles will not be near the end of their depreciation curves during our analysis period.


\subsection{Dataset Construction}
\label{sec:construction}

We merge the music consumption data from Luminate with our snapshot of the iTunes Music Store catalog to construct two datasets, one at the recording level and the other at the artist-title level. 

\subsubsection{Selecting artist-titles and recordings} \label{sec:selecting} We retrieve from Luminate the top 10,000 artist-titles with the most digital single sales in each of the years 2008 and 2009, restricted to artist-titles having rock as their core genre. We union these two ranked lists to obtain 11,790 unique artist-titles. For each of these artist-titles, we retrieve their monthly digital single sales at the recording level in 2011 and 2012. We also retrieve their monthly programmed streams, on-demand streams, and radio airplay audience impressions at the artist-title level. We exclude ISRCs that entered or exited during this period to reduce the possibility of consumption changes due to re-release promotions\footnote{\label{footnote:disturbed}For example, the recording titled ``Conflict'' by artist ``Disturbed'' is identified by 3 ISRCs \citep{mb2023-4}. The earliest of these ISRCs appeared in 2000 with the recording's first release, while a later ISRC (which we exclude) appeared in March 2010 coinciding with the recording's re-release on a ``10$^{\textrm{th}}$ Anniversary Edition'' album.}.

\begin{figure*}[!t]
  \centering
  \caption{Recording and track durations}
  \begin{subfigure}[t]{0.49\textwidth}
    \centering
    \includegraphics[width=\textwidth]{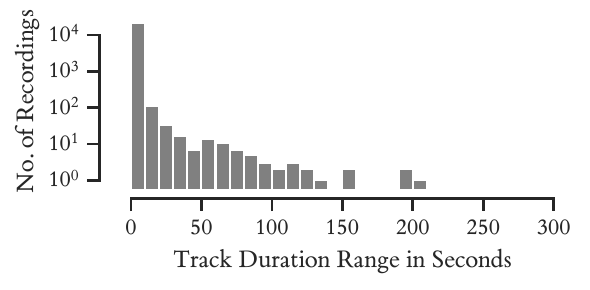}
    \caption{Track duration range distribution}
  \end{subfigure}
  \begin{subfigure}[t]{0.49\textwidth}
    \centering
    \includegraphics[width=\textwidth]{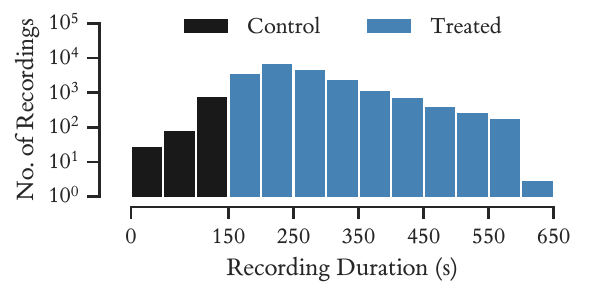}
    \caption{Recording duration distribution}
  \end{subfigure} 
  \caption*{\footnotesize\textit{Notes:} Figure \ref{fig:durationdistributions}(a) plots the distribution of the track duration ranges (the duration difference between the longest and shortest track of a recording). Figure \ref{fig:durationdistributions}(b) plots the distribution of the treated and control recording durations. Figure \ref{fig:durationdistributions}(b) excludes 2 recordings used to produce ``Space Trucking'' by artist Deep Purple with durations of 1295 and 1159 seconds.} 
  \label{fig:durationdistributions}
\end{figure*}
\subsubsection{Linking Luminate recordings with iTunes Music Store tracks} \label{sec:duration} Since Luminate reports digital single sales at the recording level, we link each recording with a representative track to determine which preview audio files to analyze.
Figure \ref{fig:durationdistributions}(a) plots the distribution of the track duration ranges (the duration difference between the longest and shortest track of a recording) and shows that 97.8\% of recordings have a track duration range below 5 seconds.  We find that the recordings with large track duration ranges are used to produce hidden tracks\footnote{Hidden tracks are prepended with long silences to separate them from the previous album track and hide them.}.

Based on Figure \ref{fig:durationdistributions}(a), we represent each recording with a single track having duration equal to the median of the recording's track durations; this avoids representing recordings with hidden tracks. If there is more than one such track, we select the track with the earliest release date. We also define the duration of a recording as the median of its track durations. Figure \ref{fig:durationdistributions}(b) plots the distribution of the resulting recording durations in our dataset.

\subsubsection{Determining recordings' treatment status}
We use the lengths of the pre- and post-policy preview audio to determine recordings' treatment status. We consider recordings as untreated (control) if the lengths of the pre- and post-policy previews of all their tracks are identical. We consider recordings as treated if the lengths of the post-policy previews of all their tracks are greater than the lengths of the pre-policy previews of all their tracks.
We exclude 31 recordings whose treatment status is ambiguous, since the preview lengths of only some of their tracks increased after the policy change (i.e. these recordings had tracks both shorter and longer than 150 seconds).

\subsubsection{Constructing the artist-title level dataset} \label{sec:artisttitle} To facilitate comparison between streams and radio airplay audience impressions reported at the artist-title level and digital single sales reported at the recording level, we construct an artist-title level dataset by selecting a single representative recording for each artist-title (we select the one with the most digital single sales in January 2010) and using the digital single sales of that recording. We also set the treatment status of each artist-title as the treatment status of its representative recording.

\begin{table*}[t]
  \small
  \centering
  \caption{Dataset summary statistics}
  \begin{tabular}[t]{lrrrrrr}
  \toprule
  \multirow{2}{*}{\textit{A: Recording Level Dataset}}
    & \multicolumn{3}{c}{Treated} 
    & \multicolumn{3}{c}{Control} \\
    \cmidrule(lr){2-4} \cmidrule(lr){5-7}
    & Mean & Median & Std. Dev. & Mean & Median & Std. Dev. \\
  \midrule
  Number of recordings & 20816 & & & 892 & & \\
  Duration in seconds & 260.9 & 243.1 & 76.6 & 127.8 & 135.3 & 23.2 \\
  Total digital single sales & 12381.4 & 1182.0 & 54241.6 & 7373.9 & 866.0 & 18906.5 \\
  \midrule
  \multirow{2}{*}{\textit{B: Artist-Title Level Dataset}}
    & \multicolumn{3}{c}{Treated} 
    & \multicolumn{3}{c}{Control} \\
    \cmidrule(lr){2-4} \cmidrule(lr){5-7}
    & Mean & Median & Std. Dev. & Mean & Median & Std. Dev. \\
  \midrule
  Number of artist-titles & 10102 & & & 426 & & \\
  Duration in seconds & 249.4 & 236.3 & 65.9 
                      & 128.7 & 135.2 & 21.3\\
  Total digital single sales
                      & 24175.2 & 6600.5 & 75933.2 
                      & 14433.1 & 5752.5 & 25403.0 \\
  Total airplay audience impressions
                      & 28 $\times 10^6$ & 0.4 $\times 10^6$ & 95 $\times 10^6$
                      & 11 $\times 10^6$ & 0.09 $\times 10^6$ & 40 $\times 10^6$ \\
  Total programmed streams
                      & 220081.5 & 5050.0  & 618462.8 
                      & 124552.6 & 1106.0  & 314290.7 \\
  Total on-demand streams
                      & 142094.8 & 21525.0  & 339550.0 
                      & 40145.2  & 2459.5   & 97856.3 \\
  \bottomrule
  \end{tabular}
  \vspace{1mm}
  \caption*{\footnotesize\textit{Notes:} Statistics based on our dataset of U.S. rock music consumption between March 1, 2010 and August 31, 2011. Panel A reports statistics at the recording level and Panel B reports statistics at the artist-title level.} 
  \label{tab:descriptive}
\end{table*}%
\begin{figure*}[t]
  \centering
  \caption{Treated and control recordings' first release years}
  \begin{subfigure}[t]{0.49\textwidth}
    \centering
    \includegraphics[width=\textwidth]{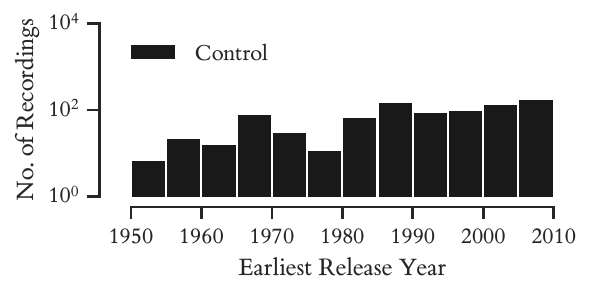}
  \end{subfigure}
  \begin{subfigure}[t]{0.49\textwidth}
    \centering
    \includegraphics[width=\textwidth]{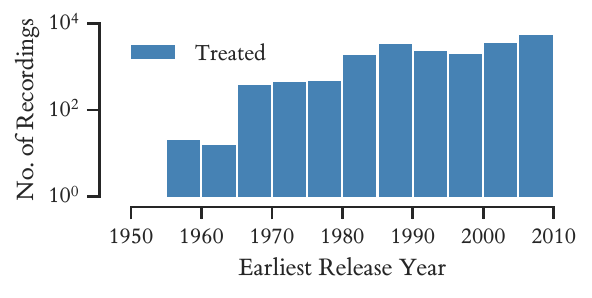}
  \end{subfigure}
  \label{fig:year}
\end{figure*}
\subsection{Descriptive Statistics}
\label{sec:descriptive}
Our sample contains 10,528 artist-titles associated with 21,708 recordings, containing 1575 hours of music and 708 hours of music excerpts. The total sales volume in our sample comprises 50-60\% of the total volume of rock music sales tracked by Luminate. Table \ref{tab:descriptive} reports summary statistics and Figure \ref{fig:year} plots the distributions of recordings' first release years.

\begin{figure*}[!t]
  \centering
  \caption{Preview start times and lengths}
  \begin{subfigure}[t]{0.495\textwidth}
    \centering
    \includegraphics[width=0.494\textwidth]{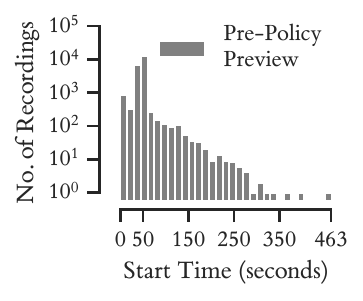}
    \includegraphics[width=0.494\textwidth]{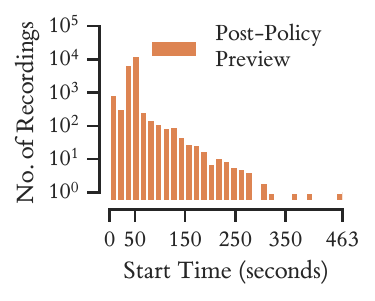}
    \caption{Pre- (left) and post-policy (right) preview start times}
  \end{subfigure}
  \begin{subfigure}[t]{0.495\textwidth}
    \centering
    \includegraphics[width=0.494\textwidth]{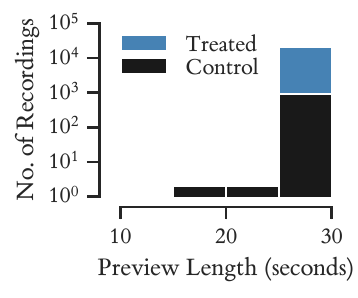}
    \includegraphics[width=0.494\textwidth]{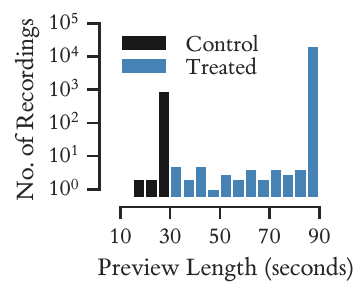}
    \caption{Pre- (left) and post-policy (right) preview lengths}
  \end{subfigure}
  \label{fig:previewalignment}
\end{figure*}
Figure \ref{fig:previewalignment}(a) plots the distributions of previews' start times\footnote{We align each preview with its corresponding recording to determine the preview start time by calculating the cross-correlation of the preview and recording audio signals and finding the earliest occurrence of the global maximum.
}. 91\% of the previews begin at 0, 30, 43, 45, or 48 seconds into the recording. Further, 99.6\% of the post-policy previews begin at exactly the same time as the pre-policy previews. The concentration of preview start times is likely due to artists being disallowed from uploading tracks to iTunes directly\footnote{Most music platforms disallow artists from uploading music directly and require uploads via approved distributors.
}, and their music distributors (or the iTunes Music Store itself) using preview start time defaults.

\begin{figure*}[!t]
  \centering
  \caption{Pre- (left) and post-policy (right) preview proportions}
  \begin{subfigure}[t]{0.495\textwidth}
    \centering
    \includegraphics[width=\textwidth]{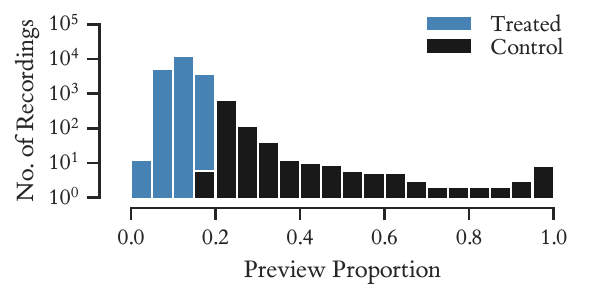}
  \end{subfigure}
  \begin{subfigure}[t]{0.495\textwidth}
    \centering
    \includegraphics[width=\textwidth]{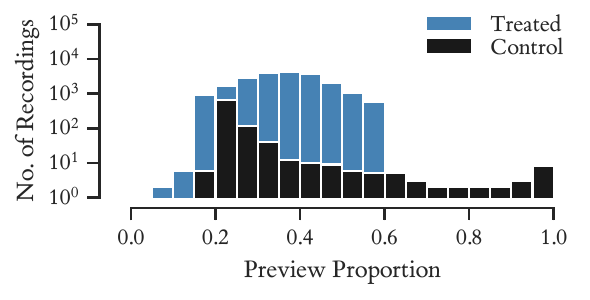}
  \end{subfigure}
  \label{fig:fraction}
  \vspace{-0.25in}
\end{figure*}
Figure \ref{fig:previewalignment}(b) plots the distributions of the preview lengths, and Figure \ref{fig:fraction} plots the distributions of the preview proportions (relative to the whole recording length) before and after the policy change. Before the policy change, treated and control recordings had previews at most 30 seconds long, with preview proportions up to 20\% for treated recordings and 100\% for control recordings. After the policy change, the lengths of treated recordings' previews increased up to 90 seconds (and proportions up to 60\%), while those of the control recordings' previews remained unchanged.

\begin{figure*}[!t]
  \centering
  \caption{Recording level U.S. digital sales volumes and trends (March 1, 2010 --- August 31, 2011)}
  \begin{subfigure}[t]{0.55\textwidth}
    \centering
    \includegraphics[width=0.49\textwidth]{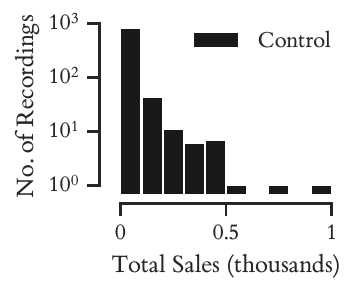}
    \includegraphics[width=0.49\textwidth]{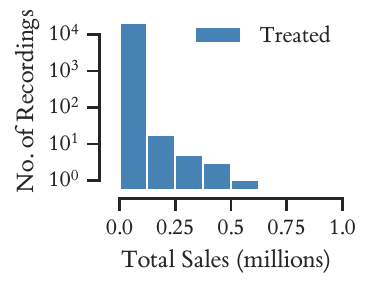}
    \caption{Total sales (volumes)}
  \end{subfigure}
  \begin{subfigure}[t]{0.44\textwidth}
    \centering
    \includegraphics[width=\textwidth]{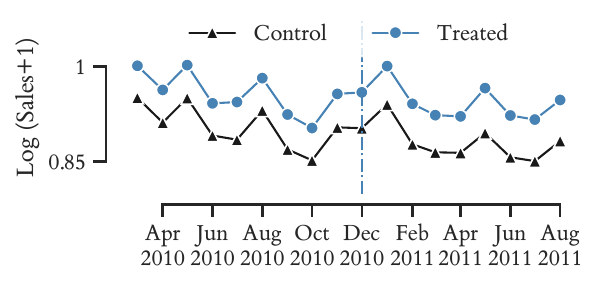}
    \caption{Monthly average log sales (volumes)}
  \end{subfigure}
  \caption*{\footnotesize\textit{Notes:} Figure \ref{fig:descriptive}(a) plots the distributions of the total digital single sales. Figure \ref{fig:descriptive}(b) plots the monthly average log digital single sales (incremented by 1). The vertical blue line indicates the policy change month. (Log) sales volumes in each plot are reported as fractions of the plot maximum to maintain data confidentiality.} 
  \label{fig:descriptive}
\end{figure*}
\begin{figure*}[t]
  \centering
  \caption{Artist-title level U.S. music consumption trends (March 1, 2010 --- August 31, 2011)}
  \begin{subfigure}[t]{0.49\textwidth}
    \centering
    \includegraphics[width=\textwidth]{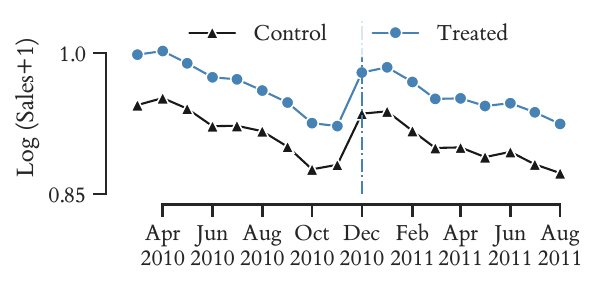}
    \caption{Monthly average log digital single sales}
  \end{subfigure}
  \begin{subfigure}[t]{0.49\textwidth}
    \centering
    \includegraphics[width=\textwidth]{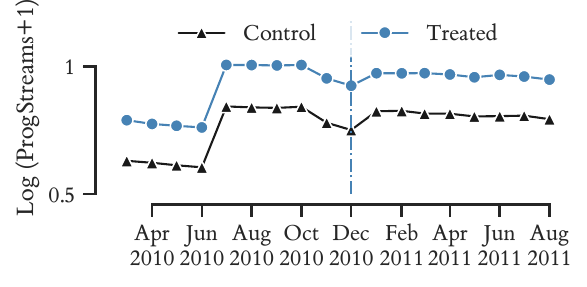}
    \caption{Monthly average log programmed streams}
  \end{subfigure}\\
  \begin{subfigure}[t]{0.49\textwidth}
    \centering
    \includegraphics[width=\textwidth]{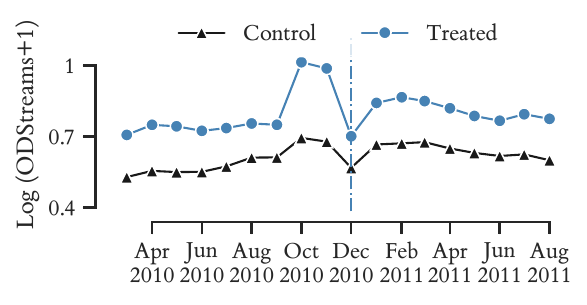}
    \caption{Monthly average log on-demand streams}
  \end{subfigure}
  \begin{subfigure}[t]{0.49\textwidth} 
    \centering 
    \includegraphics[width=\textwidth]{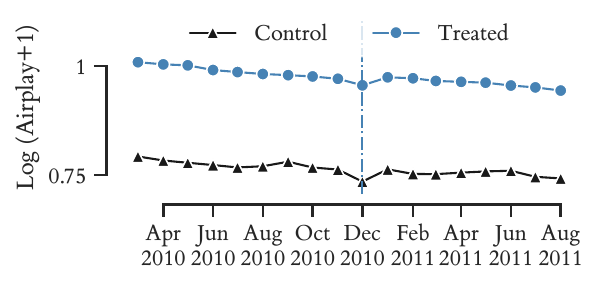}
    \caption{Monthly average log airplay audience impressions}
  \end{subfigure}
  \caption*{\footnotesize\textit{Notes:} 
  Figure \ref{fig:descriptive-songlevel} plots the monthly average (a) log digital single sales, (b) log programmed streams, (c) log on-demand streams, and (d) log airplay audience impressions for artist-titles. Consumption metrics are incremented by 1 before taking their logarithm. The vertical blue line indicates the policy change month. The log consumption volumes in each plot are reported as fractions of the plot maximum to maintain data confidentiality} 
  \label{fig:descriptive-songlevel}
\end{figure*}

Figure \ref{fig:descriptive}(a) plots the distributions of recording level total digital single sales in our dataset.
Figure \ref{fig:descriptive}(b) plots the monthly average log digital single sales (incremented by 1) of treated and control recordings, and provides model-free evidence of parallel trends in their digital single sales before the policy change.
Figure \ref{fig:descriptive-songlevel} plots the monthly average log consumption (incremented by 1) of artist-titles in our artist-title level dataset, and provides model free evidence of parallel pre-policy consumption trends at the artist-title level with the exception of on-demand streams.

We investigated possible explanations for the surge in on-demand streams in October and November 2010 in Figure \ref{fig:descriptive-songlevel}(c).
In communication with Luminate, we were informed that this anomaly is due to MySpace beginning to provide data to Luminate in October 2010, coinciding with MySpace's relaunch and promotional activity. The higher surge in streams for treated artist-titles can be attributed due to MySpace hosting a larger number of songs and music videos longer than 150 seconds.
Since the MySpace relaunch did not coincide with the iTunes policy change, we include results with on-demand streams as a consumption outcome for completeness.

\section{The Impact of the iTunes Preview Policy Change}
\label{sec:policy}

In this section, we quantify the average impact of the iTunes preview policy change on treated recordings. We restrict the treatment to be a binary indicator in this section, but relax this restriction in Section \ref{sec:information} to consider the impact of continuous changes in preview content characteristics.

\subsection{Identification and Estimation}

We begin our analyses by formalizing our estimation and inference procedure, stating our identification assumptions, and discussing potential threats to identification.

\subsubsection{Estimation and inference}
\label{sec:estimation}
Our main estimating equations take the following form:
\begin{align} 
  \textrm{log}(Y_{it} + 1) =
    \alpha + \beta D_{it} + \delta_i + \gamma_t + \epsilon_{it},
  \label{eq:twfe}
\end{align}
where $\alpha$ is a constant and $\epsilon_{it}$ is an error term. Equation \ref{eq:twfe} relates the consumption metric $Y_{it}$ logarithmically to unit $i$'s treatment status $D_{it}$ in month-year $t$ controlling for unit fixed effects $\delta_i$ and month-year fixed effects $\gamma_t$. The treatment indicator is defined as $D_{it} = \textrm{Treated}_i \times \textrm{Post}_t$, where $\textrm{Treated}_i$ indicates whether unit $i$ is treated and $\textrm{Post}_t$ indicates whether $t \geq \textrm{December 2010}$.

In our main analyses at the recording level, the consumption metric $Y_{it}$ is the digital single sales of recording $i$ in month-year $t$. In our supplementary analyses at the artist-title level, we consider four consumption metrics for each artist-title $i$ in month-year $t$: digital single sales, programmed streams, on-demand streams, and radio airplay audience impressions.

We report the OLS estimates $\hat{\beta}$ with robust standard errors clustered by recording (in our recording level analyses) or artist-title (in our artist-title level analyses) to address within-unit serial correlation over time \citep{bertrand2004much,abadie2023should}. To increase precision, we also report estimates after controlling for the number of years elapsed between month-year $t$ and the first release date of the recording $i$, denoted by $\textrm{Age}_{it}$. We include $\textrm{Age}_{it}$ as a set of fixed effects (one for each value of $\textrm{Age}_{it}$) to avoid introducing additional functional form assumptions.

\subsubsection{Identification} \label{sec:identification} To interpret $\hat{\beta}$ as a causal effect, we rely on identification assumptions that we formalize using the potential outcomes framework \citep{imbens2015causal}. Let $Y^a_{it}$ be the potential outcome of unit $i$ in month-year $t$ under treatment status $a \in \{0,1\}$. Then, we assume:
\begin{assumption}
  For all $t$: $\mathbb{E}[Y^0_{it} - Y^0_{it-1}]$ does not vary with $i$.
  \label{assumption:paralleltrends}
\end{assumption}
\begin{assumption}
  For all $i$ and $t$: $Y_{it} = (1-D_{it})Y^0_{it} + D_{it}Y^1_{it}$.
  \label{assumption:consistency}
\end{assumption}
Assumption \ref{assumption:paralleltrends} generalizes the 2-period 2-group parallel trends assumption \citep{abadie2005semiparametric} to multiple groups and periods \citep{de2020two}.
Assumption \ref{assumption:consistency} formalizes the consistency assumption that is often implicitly assumed.
Under Assumptions \ref{assumption:paralleltrends} and \ref{assumption:consistency} and assuming that all expectations exist, $\hat{\beta}$ is the average treatment effect on the treated (ATT). Since Equation \ref{eq:twfe} is log-linear, we interpret this estimate as a 100$\times \hat{\beta}\%$ percentage change.

\subsubsection{Potential threats to identification} \label{sec:threats} Assumption \ref{assumption:paralleltrends} will be violated by time-varying factors that differentially affect the consumption of treated and control units. We note a few such factors below, and show empirical support for having mitigated them in Section \ref{sec:mainresults} and Section \ref{sec:placebo}.

Taste depreciation is one such factor that may bias our estimates downwards, as discussed in Section \ref{sec:depreciation}. Promotional activity is another such factor.
When recordings are released, performed live, or promoted otherwise, their consumption surges \citep{moe2001modeling,lee2003bayesian,papies2017dynamic}. Surge magnitudes can differ for treated and control units, and hence promotional activity can introduce bias in either direction depending on the proportion of promoted units that are treated. Our dataset construction attempts to eliminate address this bias by considering only top selling artist-titles and by excluding new recording releases in 2011 and 2012.

Assumption \ref{assumption:consistency} will be violated if, for example, the digital single sales of a treated recording are mostly from the Amazon MP3 store (where it was not treated at all, since the Amazon MP3 store uses 30-second previews).
Such violations of Assumption \ref{assumption:consistency} would bias our estimates towards zero. We do not attempt to mitigate this bias and interpret our estimates as conservative.

\begin{table*}[!t]
  \small
  \centering
  \caption{Estimated effects of the iTunes preview policy change on recording level digital single sales}
  \begin{tabular}[t]{l@{\hskip 25mm}ccccc}
    \toprule
    Covariate (Indicator) & (1) & (2) & (3) & (4) & (5)\\
    \midrule
    $\textrm{Treated}_i \times \textrm{Post}_t = D_{it}$ 
      & $\phantom{-}0.042$$^{***}$
      & $\phantom{-}0.042$$^{***}$
      & $\phantom{-}0.054$$^{***}$ 
      & $\phantom{-}0.041$$^{***}$ 
      & $\phantom{-}0.041$$^{***}$ \\
      & ($0.02$)
      & ($0.02$)
      & ($0.02$)
      & ($0.02$)
      & ($0.01$)\\[0.5em]
    $\textrm{Treated}_i$
      & $\phantom{-}0.221$$^{***}$
      & --- & --- & --- &--- \\
      & ($0.08$)
      &  &  &  & \\[0.5em]
    $\textrm{Post}_t$ 
    & $-0.098$$^{***}$
    & --- & --- & --- & --- \\
    & ($0.02$)
    &  &  &  &  \\
    \midrule
    $\textrm{Age}_{it}$ fixed effects
      & \xmark & \xmark & \cmark & \xmark & \xmark \\
    Recording fixed effects $\delta_i$                
      & \xmark & \cmark & \cmark & \cmark & \cmark \\
    Month-year fixed effects $\gamma_t$          
      & \xmark & \cmark & \cmark & \cmark & \cmark\\
    No. of recordings                                
      & $21,708$ & $21,708$ & $21,708$ & $21,708$ & $21,708$\\
    No. of observations
      & $390,744$ & $390,744$ & $390,744$ & $390,744$ & $390,744$\\
    \bottomrule
  \end{tabular}
  \vspace{1mm}
  \caption*{\footnotesize\textit{Notes:} Table \ref{tab:mainresults} reports the estimated effects of the iTunes preview policy change, with the digital single sales of recording $i$ in month-year $t$ as the outcome $Y_{it}$. OLS estimates of Equation \ref{eq:twfe} are reported in columns (1), (2), and (3) with robust standard errors clustered by recording in parentheses. Heterogeneity-robust \citep{de2020two} estimates are reported in column (4) with standard errors clustered by recording in parentheses. Synthetic difference-in-differences \citep{arkhangelsky2021synthetic} estimates are reported in column (5) with jackknife standard errors in parentheses. $^{***}p<0.01; ^{**}p<0.05; ^{*}p<0.1$.} 
  \label{tab:mainresults}
  \vspace{-0.1in}
\end{table*}
\subsection{Main Results: Estimated Effects on Digital Single Sales}
\label{sec:mainresults}


In columns (1) and (2) of Table \ref{tab:mainresults}, we report OLS estimates of Equation \ref{eq:twfe} without and with recording and month-year fixed effects, respectively. In both columns, the point estimates of $\hat{\beta}$ are identical. The estimates show that longer previews increase the digital single sales of treated recordings by 4.2\% on average. In column (3), we report OLS estimates of Equation \ref{eq:twfe} after including $\textrm{Age}_{it}$ fixed effects and find a larger estimated effect of 5.4\%.

\begin{figure*}[!t]
  \centering
  \caption{Recording level digital single sales event study estimates}
  \begin{subfigure}[t]{0.49\textwidth}
    \includegraphics[width=\textwidth]{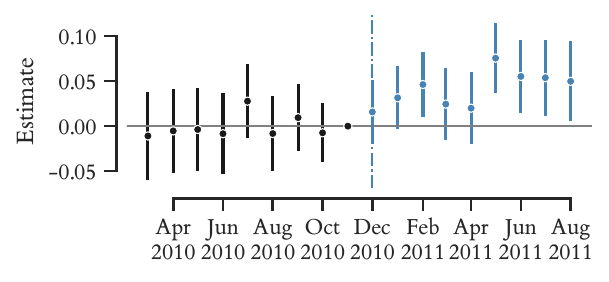}
    \caption{Not including $\textrm{Age}_{it}$ fixed effects}
  \end{subfigure}
  \begin{subfigure}[t]{0.49\textwidth}
    \includegraphics[width=\textwidth]{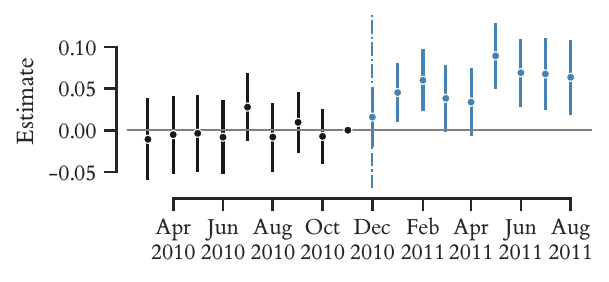}
    \caption{Including $\textrm{Age}_{it}$ fixed effects}
  \end{subfigure}
  \caption*{\footnotesize\textit{Notes:} Figure \ref{fig:eventstudy} plots the event study estimates of Equation \ref{eq:eventstudy} with $\beta_{-1}=0$ and $\textrm{Age}_{it}$ fixed effects (a) not included, and (b) included. Error bars indicate 95\% confidence intervals. The vertical blue line indicates the policy change month.}
  \label{fig:eventstudy}
\end{figure*}
We examine pre-trends in the recording level digital single sales by estimating the following event study equation \citep{freyaldenhoven2021visualization}:
\begin{align}
  \textrm{log}(Y_{it} + 1) =
  \alpha + \sum_{k=-9}^{8}\beta_k D_{it-k} + \delta_i + \gamma_t + \epsilon_{it},
  \label{eq:eventstudy}
\end{align}
where $D_{it-k}$, by definition, equals 1 if the number of months between $t$ and the policy change month equals $k$. The other terms mirror those in Equation \ref{eq:twfe}. We fix $\beta_{-1}=0$ based on Suggestion 1 in \citep{freyaldenhoven2021visualization} and interpret the remaining $\beta_k$ estimates relative to $\beta_{-1}$.
Figure \ref{fig:eventstudy}(a) plots the estimates of $\beta_k$ in Equation \ref{eq:eventstudy}, and
Figure \ref{fig:eventstudy}(b) plots these estimates after additionally controlling for $\textrm{Age}_{it}$ fixed effects. Both figures show that the pre-policy estimates of $\beta_k$ are near and statistically indistinguishable from zero, supporting Assumption \ref{assumption:paralleltrends}.

We now revisit the two identification threats discussed in Section \ref{sec:threats}. First, since our estimated effects are positive and taste depreciation biases estimates downwards, any bias due to taste depreciation will produce conservative estimates.
Second, recall that promotional activity can produce consumption surges that differ in magnitude for treated and control recordings.
If promotional activity is indeed driving all of the $\beta_k$ estimates for and after December 2010 in Figure \ref{fig:eventstudy}, such promotional activity must be (i) absent or ineffective before December 2010, and (ii) present, effective, and sustained in every month during and after December 2010. This seems implausible.
 
\subsection{Placebo Tests: Estimated Effects on Non-iTunes Music Consumption}
\label{sec:placebo}
We do not expect longer previews on iTunes to significantly affect \textit{non}-iTunes music consumption. Hence, we evaluate the presence of confounding factors coincident with the timing of the iTunes policy change that differentially affect treated and control units by estimating the impact of the iTunes policy change on three \textit{non}-iTunes consumption metrics: radio airplay audience impressions, on-demand streams, and programmed streams. Since Luminate reports these metrics at the artist-title level only, we report estimates using our artist-title level dataset described in Section \ref{sec:artisttitle}.


\begin{figure*}[!t]
  \centering
  \caption{Artist-title level event study estimates for various consumption metrics}
  \begin{subfigure}[t]{0.49\textwidth}
    \includegraphics[width=\textwidth]{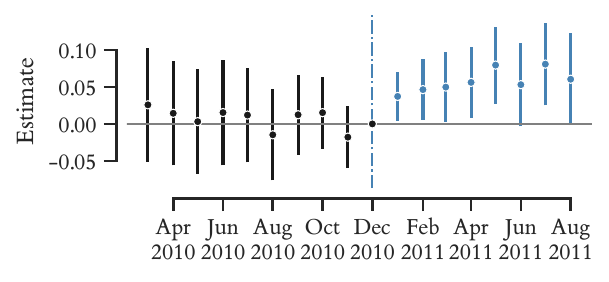}
    \caption{Digital single sales}
  \end{subfigure}
  \begin{subfigure}[t]{0.49\textwidth}
    \includegraphics[width=\textwidth]{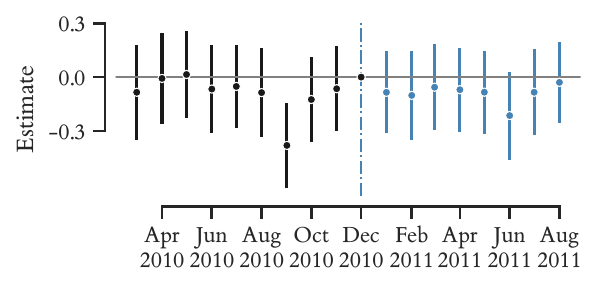}
    \caption{Airplay audience impressions}
  \end{subfigure}\\
  \begin{subfigure}[t]{0.49\textwidth}
    \includegraphics[width=\textwidth]{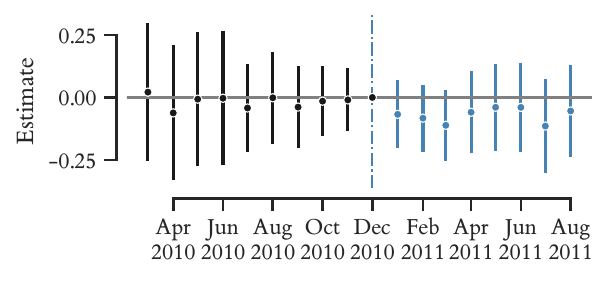}
    \caption{Programmed streams}
  \end{subfigure}
  \begin{subfigure}[t]{0.49\textwidth}
    \includegraphics[width=\textwidth]{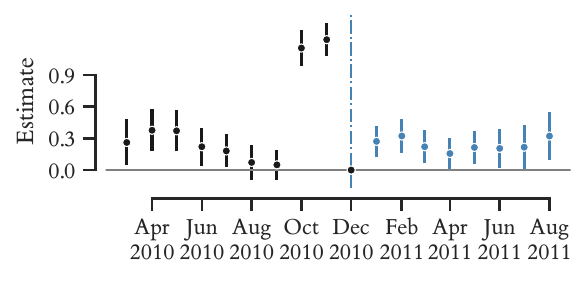}
    \caption{On-demand streams}
  \end{subfigure}
  \caption*{\footnotesize\textit{Notes:} Figure \ref{fig:placeboeventstudy} plots event study estimates of Equation \ref{eq:eventstudy} with $\beta_0=0$ and $\textrm{Age}_{it}$ fixed effects included for outcomes (a) digital single sales, (b) radio airplay audience impressions, (c) programmed streams, and (d) on-demand streams at the artist-title level. Error bars indicate 95\% confidence intervals. The vertical blue line indicates the policy change month.}
  \label{fig:placeboeventstudy}
\end{figure*}
We first examine artist-title level consumption trends by estimating Equation \ref{eq:eventstudy} for each consumption metric, with $\textrm{Age}_{it}$ included to increase precision. Since Figure \ref{fig:descriptive-songlevel}(c) shows that on-demand streams deviated anomalously in October and November 2010, we use December 2010 as the reference time period by fixing $\beta_{0}=0$. We plot the remaining $\beta_k$ estimates and their 95\% confidence intervals for each consumption metric in Figure \ref{fig:placeboeventstudy}.

Figure \ref{fig:placeboeventstudy} shows that, with the exception of on-demand streams, the pre-policy estimates of $\beta_k$ are near and jointly statistically indistinguishable from zero for all artist-title level consumption metrics. Hence, Figure \ref{fig:placeboeventstudy} supports the parallel trends assumption for the artist-title level digital single sales, radio airplay audience impressions, and programmed streams.


In Table \ref{tab:placeboresults}, we report OLS estimates $\hat{\beta}$ in Equation \ref{eq:twfe} for each artist-title level consumption metric with recording, month-year, and Age$_{it}$ fixed effects. We exclude data in October and November 2010 for the on-demand streams outcome only, based on their anomalous behavior in these months. We find a positive and statistically significant estimated effect of 4.4\% on artist-title level digital single sales. In contrast, the estimated effects of the iTunes policy change on the non-iTunes consumption metrics are all statistically insignificant.

Overall, Figure \ref{fig:placeboeventstudy} and the estimates in Table \ref{tab:placeboresults} do not support the existence of confounding factors coincident with the timing of the iTunes policy change that differentially affect the consumption of treated and control artist-titles. Had such confounders been present, we would have observed their impact on non-iTunes music consumption in addition to iTunes music consumption.

\begin{table*}[!t]
  \small
  \centering
  \caption{Estimated effects of the iTunes preview policy change on artist-title level consumption metrics}
  \begin{tabular}[t]{l@{\hskip 7.5mm}cccc}
    \toprule
    & \multicolumn{4}{c}{Dependent Variable}\\
    \cmidrule{2-5}
    \multirow{2}{*}{Covariate (Indicator)}
    & Digital
    & Airplay 
    & Programmed
    & On-demand\\
    & Single Sales
    & Audience
    & Streams
    & Streams\\
    \midrule
    $\textrm{Treated}_i \times \textrm{Post}_t = D_{it}$ 
      & $\phantom{-}0.044$ ($0.02$)$^{*}\phantom{^{**}}$
      & $\phantom{-}0.015$ ($0.06$)$\phantom{^{***}}$
      & $-0.045$ ($0.07$)$\phantom{^{***}}$
      & $-0.021$ ($0.07$)$\phantom{^{***}}$\\
    \midrule
    $\textrm{Age}_{it}$ fixed effects
      & \cmark & \cmark & \cmark & \cmark \\
    Artist-title fixed effects $\delta_i$                
      & \cmark & \cmark & \cmark & \cmark \\
    Month-year fixed effects $\gamma_t$          
      & \cmark & \cmark & \cmark & \cmark\\
    No. of artist-titles
      & $10,528$ & $10,528$ & $10,528$ & $10,528$\\
    No. of observations
      & $189,504$ & $189,504$ & $189,504$ & $189,504$\\
    \bottomrule
  \end{tabular}
  \vspace{1mm}
  \caption*{\footnotesize\textit{Notes:} Table \ref{tab:placeboresults} reports the estimated effects of the iTunes policy change on digital single sales, radio airplay audience impressions, programmed streams, and on-demand streams. The estimate for on-demand streams excludes October and November 2010 data. Robust standard errors clustered by artist-title are in parentheses. $^{***}p<0.01; ^{**}p<0.05; ^{*}p<0.1$.} 
  \label{tab:placeboresults}
\end{table*}

\subsection{Effect Heterogeneity with Recording Popularity}
\label{sec:popularity}

Motivated by our model in Section \ref{sec:theory}, we investigate information provision as a mechanism by examining how the impact of the policy change differs between unpopular songs (having less than the median sales in the year prior to the policy change) and popular songs (having greater than or equal to the median sales in the year prior to the policy change). We expect consumers to be relatively more informed about popular songs due to the extensive reviews and word-of-mouth typically available. Hence, we expect longer excerpts to have a smaller impact on popular songs.

\begin{table*}[!t]
  \small
  \centering
  \caption{Heterogeneous effects of the iTunes preview policy change on low and high popularity recordings}
  \begin{tabular}[t]{l@{\hskip 15mm}ccc}
    \toprule
    & \multicolumn{3}{c}{Subsample of Recordings}\\
    \cmidrule{2-4}
    Covariate (Indicator) & All Recordings & Unpopular Recordings & Popular Recordings\\
    \midrule
    $D_{it} = \textrm{Treated}_i \times \textrm{Post}_t$ 
      & $\phantom{-}0.097$ ($0.02$)$^{***}$
      & $\phantom{-}0.093$ ($0.02$)$^{***}$
      & $\phantom{-}0.007$ ($0.02$)$\phantom{^{***}}$ \\
    $D_{it} \times \textrm{Popular}_i$ 
      & $-0.089$ ($0.03$)$^{***}$
      & ---
      & --- \\
    $\textrm{Post}_{t}\times\textrm{Popular}_i$ 
      & $-0.050$ ($0.03$)$^{*}\phantom{^{**}}$
      & ---
      & --- \\
    \midrule
    $\textrm{Age}_{it}$ fixed effects
      & \cmark & \cmark & \cmark \\
    Recording fixed effects $\delta_i$                
      & \cmark & \cmark & \cmark\\
    Month-year fixed effects $\gamma_t$          
      & \cmark & \cmark & \cmark\\
    No. of recordings                                
      & $21,708$ & $10,852$ & $10,856$\\
    No. of observations
      & $390,744$ & $195,336$ & $195,408$\\
    \bottomrule
  \end{tabular}
  \vspace{1mm}
  \caption*{\footnotesize\textit{Notes:} Table \ref{tab:popularity} reports the estimated effects of the iTunes policy change for popular and unpopular recordings, where $\textrm{Popular}_i$ indicates whether recording $i$ had sales greater than or equal to the median in 2009. The outcome $Y_{it}$ is the digital single sales of recording $i$ in month-year $t$. OLS estimates of Equation \ref{eq:twfe} are reported with robust standard errors clustered by recording in parentheses. $^{***}p<0.01; ^{**}p<0.05; ^{*}p<0.1$.} 
  \label{tab:popularity}
\end{table*}
\begin{figure*}[!t]
  \centering
  \caption{Event study estimates in subsamples of unpopular and popular recordings}
  \begin{subfigure}[t]{0.49\textwidth}
    \centering
    \includegraphics[width=\textwidth]{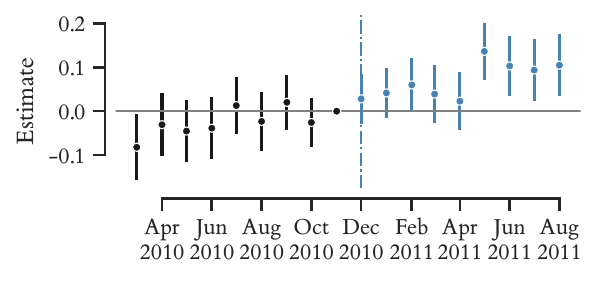}
    \caption{Unpopular recordings}
  \end{subfigure}
  \begin{subfigure}[t]{0.49\textwidth}
    \centering
    \includegraphics[width=\textwidth]{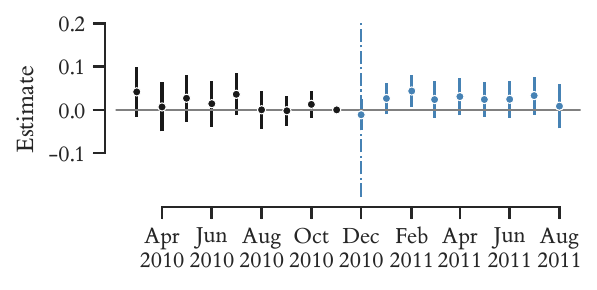}
    \caption{Popular recordings}
  \end{subfigure}
  \caption*{\footnotesize\textit{Notes:} Figure \ref{fig:eventstudypopularity} plots the event study estimates of Equation \ref{eq:eventstudy} with $\beta_{-1}=0$ for subsamples of recordings with digital single sales in 2009 that were (a) less than, and (b) greater than or equal to the median. Error bars indicate 95\% confidence intervals. The vertical blue line indicates the policy change month.}
  \label{fig:eventstudypopularity}
\end{figure*}
In Table \ref{tab:popularity}, we report OLS estimates $\hat{\beta}$ in Equation \ref{eq:twfe} after including interactions with the indicator $\textrm{Popular}_i$, which equals 1 if and only if recording $i$ had sales greater than or equal to the median in 2009. We also report OLS estimates $\hat{\beta}$ in Equation \ref{eq:twfe} in subsamples of unpopular and popular recordings. The estimates in Table \ref{tab:popularity} show that longer previews increase the digital single sales of unpopular recordings by 9.3-9.7\%, which is significantly higher (by 8.9 percentage points) than their impact on popular recordings.
In Appendix C, we further show that within the subsample of unpopular recordings, the digital single sales of recordings by unpopular artists increase more from longer previews than the digital single sales of recordings by popular artists.

In Figure \ref{fig:eventstudypopularity}, we plot the event study estimates $\hat{\beta}_k$ in Equation \ref{eq:eventstudy} estimated in subsamples of unpopular and popular recordings with $\textrm{Age}_{it}$ fixed effects included. The pre-policy estimates of $\beta_k$ in the unpopular recordings subsample are near zero, while the post-policy estimates in this subsample are positive. In contrast, both the pre- and post-policy estimates of $\beta_k$ in the popular recordings subsample are near zero. Figure \ref{fig:eventstudypopularity} thus supports the parallel trends assumption.

The disproportionate impact of longer previews on unpopular recordings, and particularly those by unpopular artists, is consistent with the information provision mechanism. For these recordings and artists, consumers likely have lower priors in terms of our model in Section \ref{sec:theory}.
\subsection{Summary and Additional Robustness Checks}

Overall, we find that longer previews increase the digital single sales of treated recordings by 4.2-5.4\% on average, and by 9.3-9.7\% for unpopular recordings. Since each sale is linked to a unique iTunes account, our estimates can be interpreted as increases in the number of unique monthly listeners of a recording (though we do not observe how many times each recording is listened to).
As such, longer previews increase consumption at least as much as being included in the Spotify Global Top 50 playlist (which increases streams by 3.3\% \citep{aguiar2018platforms}).

In columns (4) and (5) of Table \ref{tab:mainresults}, we report results from additional estimators to evaluate robustness to statistical assumptions. In column (4), we estimate an effect of 4.1\% with an estimator robust to heterogeneity \citep{de2020two}. In column (5), we estimate an effect of 4.0\% using the synthetic difference-in-differences estimator, which re-weights units and time periods to improve robustness to parallel trend violations \citep{arkhangelsky2021synthetic}.

In Appendix B, we evaluate effect heterogeneity with the recording duration (or preview proportion). We find that the impact of longer excerpts is more positive for shorter recordings (having a higher preview proportion) than for longer recordings (having a lower preview proportion). Hence, in our context, we do not find evidence of demand cannibalization. This may be explained by the maximum observed treated recording preview proportion of 60\%. Our estimates suggest that below a preview proportion of 60\%, the market expansion effect of longer previews dominates.

\clearpage

\section{What Makes an Effective Excerpt? The Role of Information}
\label{sec:information}

To better understand the information provision mechanism discussed in Section \ref{sec:popularity}, we leverage the fact that the iTunes policy change modified the audio content of the previews of treated recordings along dimensions apart from length.
Formally, the binary treatment in Section \ref{sec:policy} has multiple ``versions'' \citep{vanderweele2013causal}.
Hence, we update the treatment in this section to be non-binary and develop two new measures of musical information to use as treatments:
repetition (Section \ref{sec:repetition}) and unpredictability (Section \ref{sec:predictability}). Based on the model in Section \ref{sec:theory}, we expect these measures to reduce information in previews, and thus moderate their impact on demand.

\subsection{Musical repetition via digital audio compression}
\label{sec:repetition}

We expect musical repetition in previews to be information-reducing. Hence, if longer previews increase demand by providing information, we expect repetition to suppress the demand-enhancing effect of longer previews. We develop a unified measure of musical repetition in audio (encompassing lyrical and musical repetition) based on properties of digital audio compression.

\subsubsection{Intuition and background}
Analog (physical) sound waves from a musical performance undergo a series of transformations on the path to digital consumption. Analog sound is first sampled at a high rate (called the sample rate) by an analog-to-digital recording device, and then stored using a fixed number of bits per sample (called the bitrate).
To facilitate digital consumption, the source recordings are then compressed using algorithms called codecs (short for en\underline{cod}er-\underline{dec}oder). We derive our measure of musical repetition from the properties of codecs.

Conceptually, codecs identify waveform repetition to encode audio using fewer bits \citep{salomon2006compression}. As an illustration, consider 10 bytes of audio represented as $\mathbf{M} = $ ``aaaaabbbbb'' (assuming 1 byte per character). $\mathbf{M}$ can be stored without losing any information as $\mathbf{M}' =$ ``5a5b'', having an \textit{encoded length} of 4 bytes after applying the run-length compression algorithm \citep{robinson1967results}. In an information-theoretic sense \citep{shannon1948mathematical}, the 10 bytes in $\mathbf{M}$ contain only 4 bytes of (Shannon) information. Based on this property, we use the encoded length of the audio file of a preview as a measure of its musical repetition for a given preview length.

\subsubsection{Operationalization}
\label{sec:operationalization1} We quantify the \textit{non}-repetitiveness of a preview with audio $\mathbf{M}$ using its encoded length $H(\mathbf{M})$, where a higher $H(\mathbf{M})$ is correlated with lower musical repetition, and a lower $H(\mathbf{M})$ is correlated with higher musical repetition.
While our measure is agnostic to the choice of codec, it requires that the perceptual audio quality be identical across the audio files being compared. If the audio quality of two files differs, their encoded lengths might differ due to differences in audio quality and not differences in musical repetition. 

Importantly, our measure excludes constant bitrate (CBR) codecs. CBR-encoded lengths are content-independent and equal the audio duration multiplied by the constant encoding bitrate. CBR encoding produces predictable file sizes while sacrificing the predictability of audio quality (i.e. non-repetitive CBR-encoded music can sound worse than repetitive CBR-encoded music with the same encoded length). Variable bitrate (VBR) encoding, in contrast, leverages musical repetition for compression \citep{apple2012} to achieve a target average bitrate (which determines audio quality).

We analyze the preview files in our dataset using FFmpeg\footnote{FFmpeg is an industry-standard audio encoding and decoding software, available at \underline{\smash{https://ffmpeg.org/}}.} to confirm that all previews are VBR-encoded with the Advanced Audio Coding (AAC) codec and have identical target bitrates of 256 kilobits per second, in line with Apple's reported encoding standards \citep{itunes2012}. We exclude 64 previews that are neither 30 nor 90 seconds long. The remaining previews either increased in length by 60 seconds after the policy change, or did not increase in length at all. Using this data sample, we quantify moderation effects for a 60 second increase in the preview length. 

\begin{figure*}[!t]
  \centering
  \caption{Encoded lengths of previews before and after the policy change}
  \begin{subfigure}[t]{0.495\textwidth}
    \centering
    \includegraphics[width=\linewidth]{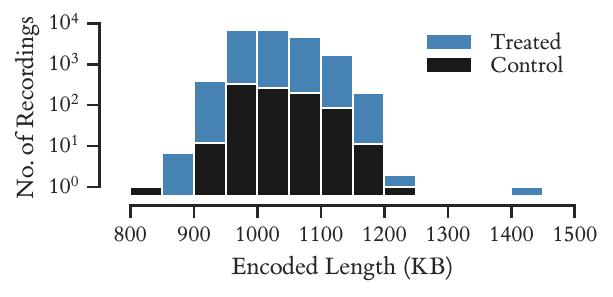}
    \caption{Before the iTunes policy change}
  \end{subfigure}
  \begin{subfigure}[t]{0.495\textwidth}
    \centering
    \includegraphics[width=\linewidth]{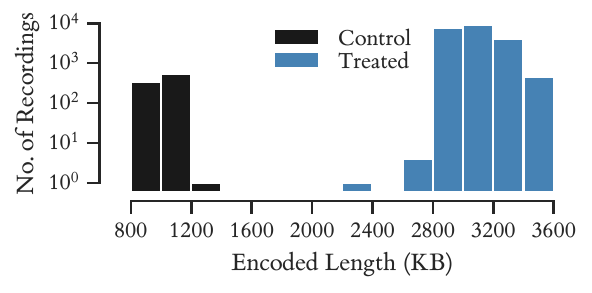}
    \caption{After the iTunes policy change}
  \end{subfigure}
  \caption*{\footnotesize\textit{Notes:} Figure \ref{fig:information2} plots the distributions of the encoded lengths (in kilobytes) of previews. 1 kilobyte (KB) = 1000 bytes.
  }
  \label{fig:information2}
\end{figure*}

We use FFmpeg to compute the number of bytes in the audio stream of each preview file, ensuring that metadata bytes are excluded. Figure \ref{fig:information2} plots the distributions of previews' encoded lengths (1KB = 1000 bytes). Before the policy change, the encoded lengths of treated and control recordings' previews are similarly distributed between 800-1500KB. After the policy change, the encoded lengths of treated recordings' previews vary between 2000-3600KB, despite being exactly 90 seconds long. We leverage this variation to quantify how the impact of having a 60 second longer preview is moderated by the preview's encoded length.

\subsubsection{Empirical strategy}
\label{sec:encodinglengthestimation}
To construct our non-binary treatment, we bin previews' encoded lengths into deciles and concatenate the indicators for each decile into a multivariate treatment.
This enables estimating the effects of each decile separately to construct a ``dose-response'' curve.

Formally, we estimate $\beta_k$ for each $k = 2, \dots, 10$ in the following equation:
\begin{align}
  \textrm{log}(Y_{it} + 1) 
  &= \alpha + \pmb{\beta}^T \pmb{H}_{it} + \delta_i + \gamma_t + \textrm{Age}_{it} + \epsilon_{it}\\
  &=
  \alpha + \sum_{k=1}^{10}\beta_k \mathbb{I}[H(\mathbf{M}_{it}) \in \textrm{Decile } k] + \delta_i + \gamma_t + \textrm{Age}_{it} + \epsilon_{it},
  \label{eq:informationtreatment}
\end{align}
where $Y_{it}$ is the digital single sales of recording $i$ in month-year $t$, $\alpha$ is a constant, $\delta_i$ is a recording fixed effect, $\gamma_t$ is a month-year fixed effect, $\textrm{Age}_{it}$ is a recording age fixed effect, and $\epsilon_{it}$ is an error term. The treatment $\pmb{H}_{it}$ is a vector of indicators $\pmb{[}\mathbb{I}[H(\mathbf{M}_{it}) \in \textrm{Decile k}]\pmb{]}_{k=1,\dots,10}$ specifying the encoded length decile of the preview $\mathbf{M}_{it}$ of recording $i$ in month-year $t$.

We report OLS estimates of the coefficients in Equation \ref{eq:informationtreatment} with robust standard errors clustered by recording. We set $\beta_1=0$ and interpret the estimate of each $\beta_k$ as the effect of being treated with a 60-second longer preview in encoded length decile $k$ relative to the effect of being treated with a 60-second longer preview in the first encoded length decile. Not all treatments are empirically observed: no recordings are treated with longer previews in the first 5 encoded length deciles. We expect the estimated effects for these deciles to be statistically indistinguishable from zero.

\begin{figure*}[!tp]
  \centering
  \caption{Estimated effects of having a 60 second longer preview in each encoded length decile}
  \begin{subfigure}[t]{0.49\textwidth}
    \centering
    \includegraphics[width=\textwidth]{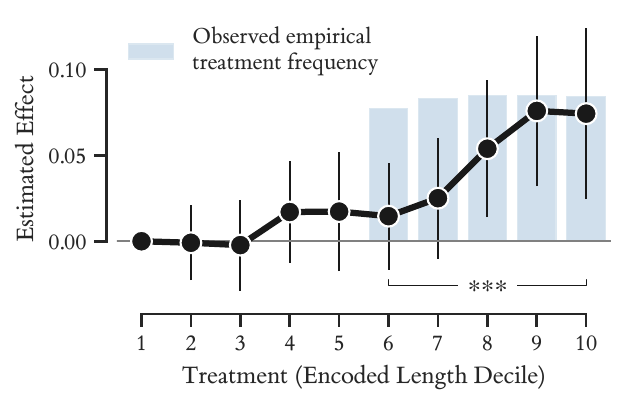}
    \caption{All recordings}
  \end{subfigure}
  \begin{subfigure}[t]{0.49\textwidth}
    \centering
    \includegraphics[width=\textwidth]{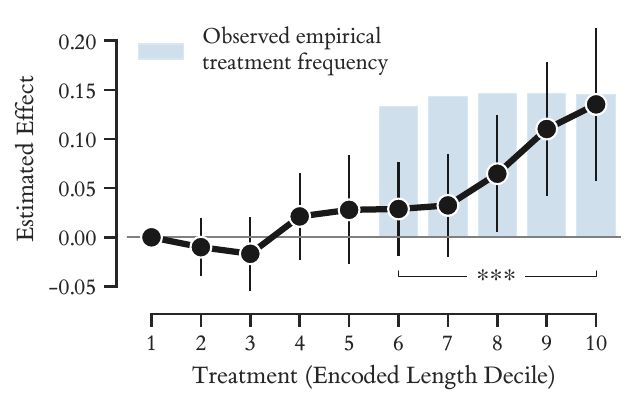}
    \caption{Unpopular recordings}
  \end{subfigure}
  \caption*{\footnotesize\textit{Notes:}  
  Figure \ref{fig:information3} plots the estimated effects of each encoded length decile on the digital single sales of treated recordings and their 95\% confidence intervals relative to the first decile (see Equation \ref{eq:informationtreatment}) for (a) all, and (b) unpopular recordings.
  }
  \label{fig:information3}
  \vspace{-4.5mm}
\end{figure*}
\subsubsection{Results}
We report the estimates of $\beta_k$ for each $k=2,\dots,10$ and their 95\% confidence intervals for all recordings in Figure \ref{fig:information3}(a), and for the subsample of unpopular recordings (as defined in Section \ref{sec:popularity}) in Figure \ref{fig:information3}(b).
We also overlay the frequency of empirically observed treatments in each decile to delineate the treatment effects that are statistically estimable.
We find that that the impact of having a 60 second longer preview varies with the preview's encoded length.

Having a 60 second longer preview in the 10$^\textrm{th}$ encoded length decile (3206-3529KB) has a statistically significant 6 percentage point higher impact on digital single sales than having one in the 6$^\textrm{th}$ encoded length decile (1139-2959KB). For unpopular recordings, this difference is larger at 10.6 percentage points. Our results indicate that the demand-enhancing effect of longer previews is suppressed when their musical repetition is high (i.e. when their encoded length is low).

\subsubsection{Discussion}
We expect musical repetition to reduce information. In the context of the model in Section \ref{sec:theory}, our results support the information provision mechanism: longer previews are more demand-enhancing when they contain more information by having less repetition.

Figure \ref{fig:information3} might, alternatively, be explained by a distaste for repetitive rock music.
Such distaste would be observable as a positive correlation between the encoded lengths of full recordings and their digital sales. Hence, we calculate the correlation between the encoded lengths of recordings and their pre-policy total digital single sales. Since recordings have different durations, we normalize recordings' encoded lengths by their durations. We find that the correlation is weak (Pearson's $r=0.01$). Hence, we do not find empirical evidence of a distaste for repetitive rock music.

\citep{nunes2015power} note that extreme musical non-repetition should be undesirable based on optimal complexity theory \citep{berlyne1971aesthetics}. However, extreme non-repetition using our proposed measure is achievable only by music resembling white noise. Since commercially viable music in the Western hemisphere is largely tonal (repetitive by design) \citep{aiello1994musical}, we expect that Figure \ref{fig:information3} is right-censored, and would otherwise reveal an ``inverted-U'' shape.
\subsection{Musical unpredictability via generative machine learning models of music}
\label{sec:predictability}

While non-repetition can increase the amount of distinct information available for consumers to learn from, distinct information may provide little marginal value if it is predictable (or expected) by consumers. Hence, if longer previews increase demand by
providing information, we expect predictability to suppress the demand-enhancing effect of longer previews.
To evaluate this, we measure musical unpredictability using generative machine learning models of music.

\subsubsection{Intuition and background}
Generative machine learning models of music are trained on tens of thousands of hours of music to generate music both unconditionally and conditionally given an audio clip \citep{dhariwal2020jukebox,copet2023simple}. We rely on \textit{autoregressive} generative music models for their computational benefits discussed later in this section.

Conceptually, autoregressive generative music models are a collection of conditional distributions over discrete \textit{tokens}, typically parameterized using transformers \citep{vaswani2017attention}. To generate the next token $s_{N+1}$ in a sequence $[s_1, \dots, s_N]$, these models sample from the conditional distribution $P(\cdot |s_1, \dots, s_N; \hat{\Theta})$, where $\hat{\Theta}$ are the trained model parameters. After generation, tokens are transformed into audio using a \textit{neural codec}.
The conditional probabilities from an autoregressive generative music model can be used to assign a probability to any token sequence.

We approximate consumers' predictive processing of music with the sequential sampling process of autoregressive generative music models. Consider a token sequence $[s_1, \dots, s_N]$. The probability assigned to this sequence by an autoregressive music model with parameters $\hat{\Theta}$ is given by the following product of conditional probabilities:
\begin{align}
  P(s_1, \dots, s_N; \hat{\Theta}) = P(s_1; \hat{\Theta})P(s_2|s_1; \hat{\Theta})P(s_3|s_1,s_2; \hat{\Theta}) \dots P(s_N|s_1,\dots,s_{N-1}; \hat{\Theta})
  \label{eq:totalprob}
\end{align}
Each conditional probability $P(s_i|s_1,\dots,s_{i-1};\hat{\Theta})$ can be interpreted as the accuracy of the model (i.e. the ``artificially intelligent'' consumer) in predicting the next token $s_i$ given the prior tokens $[s_1,\dots,s_{i-1}]$.
Each conditional distribution $P(\cdot|s_1,\dots,s_{i-1};\hat{\Theta})$ is obtained using a \textit{forward pass} of the sequence $[s_1,\dots,s_{i-1}]$ through the model. The conditional probabilities can thus be computed in parallel and without multiple sampling iterations. This computational advantage is absent in non-autoregressive models (eg. diffusion-based models \citep{schneider2023mo}).

\subsubsection{Operationalization}
\label{sec:perplexity}
We use the ``melody'' pretrained checkpoint of the MusicGen
generative music model \citep{copet2023simple}. MusicGen is trained on about 20,000 hours of music, of which about 10,000 hours are from licensed music internal to Meta and the remaining are from stock music on Pond5 and ShutterStock.
We use the EnCodec neural codec \citep{defossez2022high}, which was also used when training MusicGen, to convert audio to discrete tokens and vice versa.

We measure the unpredictability of a preview as the log-perplexity assigned to its EnCodec-tokenized audio by the MusicGen model. The log-perplexity of a sequence is the negative logarithm of its probability, given by Equation \ref{eq:totalprob} \citep{chen1998evaluation}. Hence, if $[s_1, \dots, s_N]$ is the tokenized audio $\mathbf{M}$ and $\hat{\Theta}$ are the MusicGen model parameters, our measure is given by:
\begin{align}
  \textrm{log-perplexity}(\mathbf{M}) = -\textrm{log}[P(s_1, \dots, s_N;\hat{\Theta})] 
  = -\sum_{i=1}^N \textrm{log}[P(s_i|s_1,\dots,s_{i-1};\hat{\Theta})]
  \label{eq:perplexity}
\end{align}
Equation \ref{eq:perplexity} quantifies the average ``surprise'' of an artificially intelligent music consumer listening to the preview with audio $\mathbf{M}$. Taking the logarithm of the probability product in Equation \ref{eq:perplexity} avoids numerical error propagation from multiplying many small probabilities. 

\begin{figure*}[!t]
  \centering
  \caption{Log-perplexities of previews before and after the policy change}
  \begin{subfigure}[t]{\textwidth}
    \centering
    \includegraphics[width=0.37\linewidth]{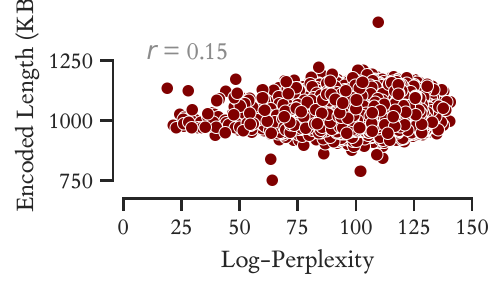}
    \includegraphics[width=0.35\linewidth]{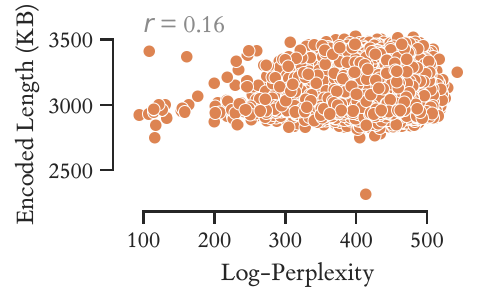}
    \caption{Encoded lengths (kilobytes) and log-perplexities of 30 (left) and 90 (right) second previews (Pearson's $r$ in grey text)}
  \end{subfigure}\\
  \begin{subfigure}[t]{0.32\textwidth}
    \centering
    \includegraphics[width=\linewidth]{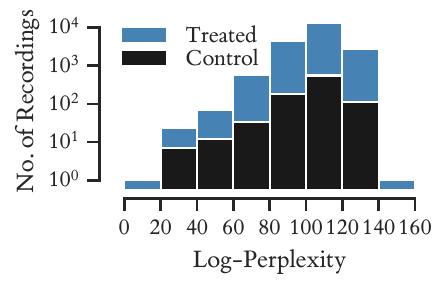}
    \caption{Before the iTunes policy change}
  \end{subfigure}
  \begin{subfigure}[t]{0.58\textwidth}
    \centering
    \includegraphics[width=0.475\linewidth]{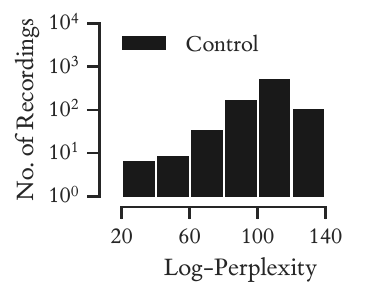}
    \includegraphics[width=0.475\linewidth]{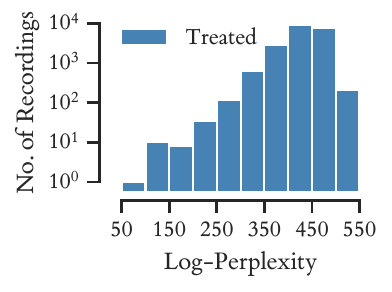}
    \caption{After the iTunes policy change}
  \end{subfigure}
  \caption*{\footnotesize\textit{Notes:} Figure \ref{fig:perplexity_distribution}(a) visualizes the correlation between the encoded lengths and log-perplexities of 30 and 90 second previews. Figures \ref{fig:perplexity_distribution}(b)-(c) plot the distributions of the log-perplexities before and after the iTunes policy change.
  }
  \label{fig:perplexity_distribution}
\end{figure*}
\subsubsection{Empirical strategy}

We use the same sample of previews as in Section \ref{sec:operationalization1}.
We find that MusicGen does poorly at generating vocals and is consistently ``surprised'' by vocals in music. Hence, we use a transformer-based stem separation model \citep{rouard2022hybrid} to remove the vocals from each preview audio. We tokenize the non-vocal preview audios with EnCodec, and reverse-engineer
the MusicGen source code to output raw conditional probabilities for any input sequence instead of generating an audio continuation. For any token $s_i$ preceded by tokens $[s_1, \dots, s_{i-1}]$, MusicGen outputs 4 conditional probabilities simultaneously\footnote{This is because the EnCodec tokenizer transforms each 0.02 second long audio chunk into 4 concurrent tokens.} after a forward pass of $[s_1, \dots, s_{i-1}]$. We take the average of these probabilities to use as $P(s_i|s_1, \dots, s_{i-1}; \hat{\Theta})$.

Figure \ref{fig:perplexity_distribution}(a) visualizes the correlation between the encoding lengths and log-perplexities of previews. The two measures are positively but weakly correlated (Pearson's $r=0.15$ for 30 second and $r=0.16$ for 90 second previews), suggesting that they capture different types of musical information.
Figures \ref{fig:perplexity_distribution}(b)-(c) plot the distributions of previews' log-perplexities. We leverage the variation in previews' log-perplexities at each preview length to quantify how the impact of having a 60 second longer preview is moderated by the preview's log-perplexity.

Similar to Section \ref{sec:encodinglengthestimation}, we bin previews' log-perplexities into deciles and concatenate the indicators for each decile into a multivariate treatment. We use this treatment in Equation \ref{eq:informationtreatment} and report the OLS estimates for each $\beta_k$ for $k=2,\dots,10$ with robust standard errors clustered by recording. We set $\beta_1$ = 0 and interpret the estimate of each $\beta_k$ as relative to $\beta_1$.

\begin{figure*}[!tp]
  \centering
  \caption{Estimated effects of having a 60 second longer preview in each log-perplexity decile}
  \begin{subfigure}[t]{0.49\textwidth}
    \centering
    \includegraphics[width=\textwidth]{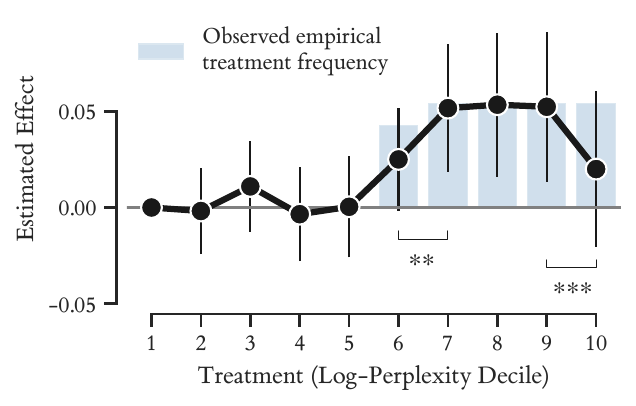}
    \caption{All recordings}
  \end{subfigure}
  \begin{subfigure}[t]{0.49\textwidth}
    \centering
    \includegraphics[width=\textwidth]{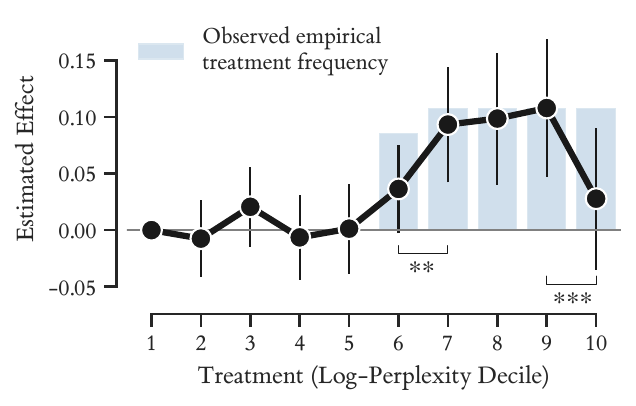}
    \caption{Unpopular recordings}
  \end{subfigure}
  \caption*{\footnotesize\textit{Notes:}  
  Figure \ref{fig:perplexity_effect} plots the estimated effects of each log-perplexity decile on the digital single sales of treated recordings and their 95\% confidence intervals relative to the first decile (see Equation \ref{eq:informationtreatment}) for (a) all, and (b) unpopular recordings.
  }
  \label{fig:perplexity_effect}
\end{figure*}
\subsubsection{Results}
We report the estimates of $\beta_k$ for each $k=2, \dots, 10$ and their 95\% conﬁdence intervals for all recordings in Figure \ref{fig:perplexity_effect}(a), and for the subsample of unpopular recordings (as deﬁned in Section \ref{sec:popularity}) in Figure \ref{fig:perplexity_effect}(b). We also overlay the frequency of empirically observed treatments in each decile to delineate the treatment effects that are statistically estimable.

We find that the moderating effect of log-perplexity follows an ``inverted-U'' shape. Moving from the 6$^{\textrm{th}}$ to the 7$^{\textrm{th}}$ log-perplexity decile is associated with a statistically significant 2.7 percentage point higher impact, and moving from the 9$^{\textrm{th}}$ to the 10$^{\textrm{th}}$ log-perplexity decile is associated with a statistically significant 3.2 percentage point lower impact. For unpopular recordings, the effect change from the 6$^{\textrm{th}}$ to the 7$^{\textrm{th}}$ log-perplexity decile is +5.7 percentage points, and the effect change from the 9$^{\textrm{th}}$ to the 10$^{\textrm{th}}$ log-perplexity decile is -8 percentage points.
Our results indicate that the demand-enhancing effect of longer previews is suppressed when they are too predictable (i.e. when their log-perplexity is too low) and too unpredictable (i.e. when their log-perplexity is too high).

\subsubsection{Discussion}
Our results are consistent with the explanation that longer previews are more demand-enhancing when they have more marginal information by being less predictable. The lower impact of longer previews when they are too unpredictable can be explained in two ways. In terms of the model in Section \ref{sec:theory}, unpredictable previews are less informative since the information they contain is difficult to extrapolate to the full song; this suppresses their demand-enhancing effect. Behaviorally, the ``inverted-U'' shape in Figure \ref{fig:perplexity_effect} can be explained by a psychobiological liking for moderate levels of musical surprise \citep{pearce2012auditory}. Hence, there is potentially a \textit{direct} negative effect of unpredictable previews on consumption, independent of their information.



\section{Conclusion}
\label{sec:conclusion}

We show that longer excerpts increase songs' unique monthly listeners by 5.4\% on average, by 9.7\% for less popular songs, and by 11.1\% for less popular songs by less popular artists. By measuring musical repetition and unpredictability, we also show empirical support for information provision as a mechanism. More broadly, we provide empirical evidence that excerpts of media products can provide consumers information, that consumers \textit{use} this information to explore emerging products and creators, and that established products and creators are not significantly affected.

\textbf{Implications for contemporary media platforms.} On contemporary media platforms such as Spotify (without excerpts), consumers ``skip'' content they consider a poor match. Skipping after previewing just the content start is pervasive: 25\% of all songs streamed on Spotify are skipped in the first 5 seconds, and 50\% are not fully listened to \citep{montecchio2020skipping}.
Our findings suggest that media platforms can use informative excerpts tailored to the content instead of the (possibly uninformative) content start to improve content discovery\footnote{Spotify's ``TikTok-ified'' home screen redesign in 2023 is an early step in this direction.}.
Our findings also support short-form media platforms adopting longer excerpts to improve consumer-content matching efficiency, but warn of this adoption driving consumers off-platform to consume long-form content.

\textbf{Implications for media marketers.}
Our findings encourage using longer promotional clips on short-form social media for emerging content in particular (noting our lack of observed cannibalization below a clip proportion of 60\%), as long as the content of the clips is not too repetitive, too predictable, or too unpredictable. For music marketers in particular, our measures of musical repetition and unpredictability can inform excerpt selection.

\textbf{Limitations.} Our work has three main limitations. First, we study consumption in online music stores, which impose a relatively high cost on exploration \citep{datta2018changing}. Excerpts may affect consumption differently when exploration costs are low. Second, we focus on rock music that depreciates slowly, and anticipate different findings in genres with a shorter shelf-life. Third, the frictions lowering the cannibalization risk in our context differ from those in contemporary social media platforms. Hence, cannibalization may be a larger concern on platforms like TikTok, where viewers can bookmark clips they like to re-listen to without needing to click-through to Spotify.

\setlength{\bibsep}{5pt plus 0.3ex}
\begin{small}
\singlespacing
\bibliographystyle{emaad}
\bibliography{scibib}
\end{small}

\clearpage

\section*{Appendix A: Apple's Note to Label Representatives}

\begin{small}
  \begin{quote}
  Dear Label Representative:

  We are pleased to let you know that we are preparing to increase the length of music previews from 30 seconds to 90 seconds on the iTunes Store in the United States. We believe that giving potential customers more time to listen to your music will lead to more purchases.
  
  All you have to do is continue making your content available on the iTunes Store, which will confirm your acceptance to the following terms.
  
  You agree that this letter modifies our U.S. Digital Music Download Sales Agreement so that “Clips” for songs longer than 2 minutes and 30 seconds may be up to 90 seconds long (``Clips'' for shorter songs will stay at 30 seconds); and you agree to license, or pass through to Apple, gratis mechanical rights for 90-second ``Clips'' embodying the entirety of compositions owned or controlled, in whole or in part, by you or your affiliates. Further, you represent that you have the authority to enter into this letter agreement for 90-second ``Clips''.

  Thank you,

  The iTunes Store Team

  Apple Inc.
\end{quote}
\end{small}

\section*{Appendix B: Effect Heterogeneity with Recording Duration}

We examine the extent of cannibalization in our context by quantifying whether longer previews differentially affect short treated recordings (shorter than the median duration) and long treated recordings (equal to or longer than the median duration) relative to all control recordings. Short recordings will have higher preview proportions and a potentially higher cannibalization risk.

\begin{table*}[!t]
  \small
  \centering
  \caption{Heterogeneous effects of the iTunes preview policy change on short and long treated recordings}
  \begin{tabular}[t]{l@{\hskip 25mm}cc}
    \toprule
    & \multicolumn{2}{c}{Subsample of Recordings (includes all control recordings)}\\
    \cmidrule{2-3}
    Covariate (Indicator)
      & Short Treated Recordings & Long Treated Recordings\\
    \midrule
    $D_{it} = \textrm{Treated}_i \times \textrm{Post}_t$ 
      & $\phantom{-}0.066$ ($0.02$)$^{***}$
      & $\phantom{-}0.039$ ($0.02$)$^{**}\phantom{^{*}}$\\
    \midrule
    $\textrm{Age}_{it}$ fixed effects
      & \cmark & \cmark \\
    Recording fixed effects $\delta_i$                
      & \cmark & \cmark \\
    Month-year fixed effects $\gamma_t$          
      & \cmark & \cmark \\
    No. of recordings                                
      & $11,300$ & $11,299$ \\
    No. of observations
      & $203,400$ & $203,382$ \\
    \bottomrule
  \end{tabular}
  \vspace{1mm}
  \caption*{\footnotesize\textit{Notes:} Table \ref{tab:duration} reports the estimated effects of the iTunes policy change in subsamples of short and long treated recordings with all control recordings. The outcome $Y_{it}$ is the digital single sales of recording $i$ in month-year $t$. OLS estimates of Eq. \ref{eq:twfe} are reported with robust standard errors clustered by recording in parentheses. $^{***}p<0.01; ^{**}p<0.05; ^{*}p<0.1$.} 
  \label{tab:duration}
\end{table*}

\begin{figure*}[!t]
  \centering
  \caption{Event study estimates in subsamples of short and long recordings}
  \begin{subfigure}[t]{0.49\textwidth}
    \centering
    \includegraphics[width=\textwidth]{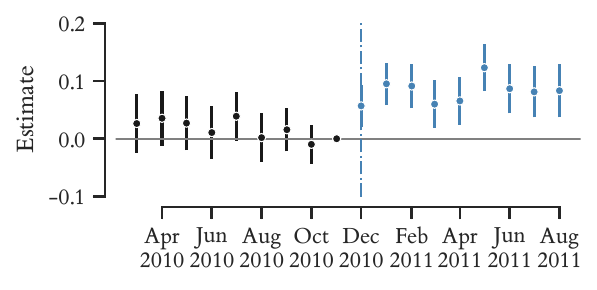}
    \caption{Short recordings}
  \end{subfigure}
  \begin{subfigure}[t]{0.49\textwidth}
    \centering
    \includegraphics[width=\textwidth]{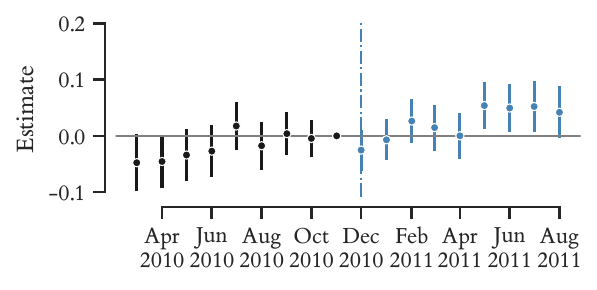}
    \caption{Long recordings}
  \end{subfigure}
  \caption*{\footnotesize\textit{Notes:} Figure \ref{fig:eventstudyduration} plots the event study estimates of Equation \ref{eq:eventstudy} with $\beta_{-1}=0$ for subsamples of (a) short, and (b) long recordings. Error bars indicate 95\% confidence intervals. The vertical blue line indicates the policy change month.}
  \label{fig:eventstudyduration}
\end{figure*}
In Table \ref{tab:duration}, we report OLS estimates $\hat{\beta}$ in Equation \ref{eq:twfe} in subsamples of short and long treated recordings, including all control recordings. The supporting event studies are in Figure \ref{fig:eventstudyduration}.
We find that longer previews increase the digital single sales of short treated recordings by 6.6\%, and that of long treated recordings by 3.9\%. Our estimates suggest that the information provided by higher preview proportions dominates cannibalization. Cannibalization in our context may be limited by the maximum observed preview proportion being 60\%, and by frictions such as the inability to listen to previews on external media devices, or to add previews to personal playlists. 

\section*{Appendix C: Effect Heterogeneity with Artist Popularity}

Similar to our analyses in Section \ref{sec:popularity}, we examine how the impact of the iTunes policy change differs for unpopular recordings by unpopular artists (with less than the median average sales over all their recordings in the year prior to the policy change) and popular artists (with greater than or equal to the median average sales over all their recordings in the year prior to the policy change).

In Table \ref{tab:artistpopularity}, we report OLS estimates $\hat{\beta}$ in Equation \ref{eq:twfe} in the subsample of unpopular recordings after including interactions with the indicator $\textrm{Popular}_i$, which equals 1 if and only if artist $i$'s average sales over all their recordings was greater than or equal to the median in 2009. We also report OLS estimates $\hat{\beta}$ in Equation \ref{eq:twfe} in subsamples of unpopular recordings by unpopular and popular artists. The estimates in Table \ref{tab:artistpopularity} show that longer previews increase the digital single sales of unpopular recordings by unpopular artists by 11.1-12.5\%, which is significantly higher than their impact on unpopular recordings by popular artists.

\begin{figure*}[!t]
  \centering
  \caption{Event study estimates in subsamples of unpopular recordings by unpopular and popular artists}
  \begin{subfigure}[t]{0.49\textwidth}
    \centering
    \includegraphics[width=\textwidth]{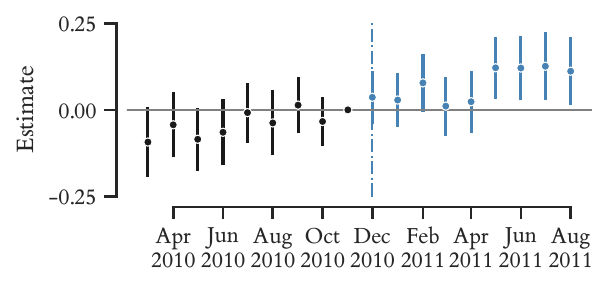}
    \caption{Unpopular artists' unpopular recordings}
  \end{subfigure}
  \begin{subfigure}[t]{0.49\textwidth}
    \centering
    \includegraphics[width=\textwidth]{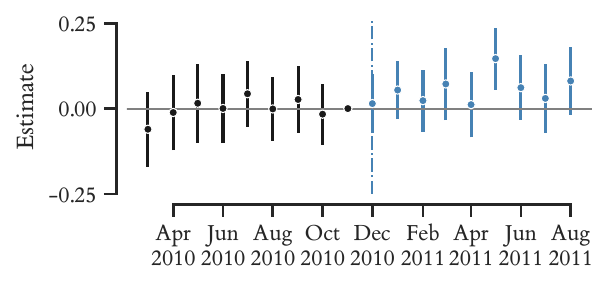}
    \caption{Popular artists' unpopular recordings}
  \end{subfigure}
  \caption*{\footnotesize\textit{Notes:} Figure \ref{fig:eventstudyartistpopularity} plots the event study estimates of Equation \ref{eq:eventstudy} with $\beta_{-1}=0$ in subsamples of unpopular recordings by artists having average digital single sales over all their recordings in 2009 (a) less than, and (b) greater than or equal to the median. Error bars indicate 95\% confidence intervals. The vertical blue line indicates the policy change month.}
  \label{fig:eventstudyartistpopularity}
\end{figure*}

In Figure \ref{fig:eventstudyartistpopularity}, we plot the event study estimates $\hat{\beta}_k$ in Equation \ref{eq:eventstudy} estimated in subsamples of unpopular recordings by unpopular and popular artists with $\textrm{Age}_{it}$ fixed effects included. The pre-policy estimates in Figure \ref{fig:eventstudyartistpopularity} support the parallel trends assumption.

\begin{table*}[!t]
  \small
  \centering
  \caption{Heterogeneous effects of the iTunes preview policy change for low and high popularity artists}
  \begin{tabular}[t]{l@{\hskip 30mm}ccc}
    \toprule
    & \multicolumn{3}{c}{Subsample of Unpopular Recordings}\\
    \cmidrule{2-4}
    Covariate (Indicator) & All Artists & Unpopular Artists & Popular Artists\\
    \midrule
    $D_{it} = \textrm{Treated}_i \times \textrm{Post}_t$ 
      & $\phantom{-}0.125$ ($0.04$)$^{**}\phantom{^{*}}$
      & $\phantom{-}0.111$ ($0.04$)$^{***}$
      & $\phantom{-}0.054$ ($0.03$)$^{*}\phantom{^{**}}$ \\
    $D_{it} \times \textrm{Popular}_i$ 
      & $-0.087$ ($0.05$)$^{* }\phantom{^{**}}$
      & ---
      & --- \\
    $\textrm{Post}_{t}\times\textrm{Popular}_i$ 
      & $-0.039$ ($0.05$)$\phantom{^{***}}$
      & ---
      & --- \\
    \midrule
    $\textrm{Age}_{it}$ fixed effects
      & \cmark & \cmark & \cmark \\
    Recording fixed effects $\delta_i$                
      & \cmark & \cmark & \cmark\\
    Month-year fixed effects $\gamma_t$          
      & \cmark & \cmark & \cmark\\
    No. of  recordings                         
      & $10,852$ & $6,677$ & $4,175$\\
    No. of observations
      & $195,336$ & $120,186$ & $75,150$\\
    \bottomrule
  \end{tabular}
  \vspace{1mm}
  \caption*{\footnotesize\textit{Notes:} Table \ref{tab:artistpopularity} reports the estimated effects of the iTunes policy change for unpopular recordings (having less than the median sales in 2009) by unpopular and popular artists, where $\textrm{Popular}_i$ indicates whether the average sales of the recordings by artist $i$ in 2009 were greater than or equal to the median. The outcome $Y_{it}$ is the digital single sales of recording $i$ in month-year $t$. OLS estimates of Equation \ref{eq:twfe} are reported with robust standard errors clustered by recording in parentheses. $^{***}p<0.01; ^{**}p<0.05; ^{*}p<0.1$.} 
  \label{tab:artistpopularity}
\end{table*}




\end{document}